\documentclass[apj]{emulateapj}

\usepackage{natbib}
\usepackage{amssymb,amsmath}
\usepackage{graphicx}
\newcommand{\degree}{$^{\circ}$~}
\newcommand{\degreee}{$^{\circ}$}

\newcommand{\kmsec}{km~s$^{-1}$}
\newcommand{\kmsecc}{km~s$^{-1}$ }
\newcommand{\ST}{\emph{STEREO} }
\newcommand{\STA}{\emph{STEREO-A} }
\newcommand{\STB}{\emph{STEREO-B} }

\newcommand{\WIN}{\emph{Wind} }

\slugcomment{\apj, accepted 2014 April 11}

\shorttitle{Connecting CME observations from the Sun to 1 AU}
\shortauthors{M\"ostl et al.}

\begin{document}

\title{Connecting speeds, directions and arrival times of 22 coronal mass ejections from the Sun to 1 AU}


\author{C. M\"ostl\altaffilmark{1,2,3},  K. Amla\altaffilmark{4}, J. R. Hall\altaffilmark{4}, P. C. Liewer\altaffilmark{4}, E. M. De Jong\altaffilmark{4},
R. C. Colaninno\altaffilmark{5}, A. M. Veronig\altaffilmark{1}, T. Rollett\altaffilmark{1}, M. Temmer\altaffilmark{1}, V. Peinhart\altaffilmark{1}, J. A. Davies\altaffilmark{6}, N. Lugaz\altaffilmark{7},  Y. D. Liu\altaffilmark{8}, C.J. Farrugia\altaffilmark{7},  J. G.  Luhmann\altaffilmark{2}, B. Vr\v{s}nak\altaffilmark{9}, R. A. Harrison\altaffilmark{6}, A. B. Galvin\altaffilmark{7}}
\affil{\altaffilmark{1}Kanzelh\"ohe Observatory-IGAM, Institute of Physics, University of Graz, Austria \url{christian.moestl@uni-graz.at}}
\affil{\altaffilmark{2}Space Science Laboratory, University of California, Berkeley, CA, USA}
\affil{\altaffilmark{3}Space Research Institute, Austrian Academy of Sciences, Graz, Austria}
\affil{\altaffilmark{4}Jet Propulsion Laboratory, California Institute of Technology, Pasadena, CA, USA}
\affil{\altaffilmark{5}Space Sciences Division, Naval Research Laboratory, Washington, DC, USA}
\affil{\altaffilmark{6}RAL Space, Harwell Oxford, Didcot, UK}
\affil{\altaffilmark{7}Space Science Center and Department of Physics, University of New Hampshire, Durham, NH, USA }
\affil{\altaffilmark{8}State Key Laboratory of Space Weather, National Space Science Center, Chinese Academy of Sciences, Beijing, China}
\affil{\altaffilmark{9}Hvar Observatory, Faculty of Geodesy, University of Zagreb, Ka\v{c}i\'{c}eva 26, HR-10000, Zagreb, Croatia.}.


\begin{abstract}
Forecasting the in situ properties of coronal mass ejections (CMEs) from remote images is expected to strongly enhance predictions of space weather, and is of general interest for studying the interaction of CMEs with planetary environments. We study the feasibility of using a single heliospheric imager (HI) instrument, imaging the solar wind density from the Sun to 1 AU, for connecting remote images to in situ observations of CMEs. We compare the predictions of speed and arrival time for 22 CMEs (in 2008-2012) to the corresponding interplanetary coronal mass ejection (ICME) parameters at in situ observatories (\emph{STEREO PLASTIC/IMPACT}, \emph{Wind SWE/MFI}). The list consists of front- and backsided, slow and fast CMEs (up to 2700 \kmsec). We track the CMEs to $34.9 \pm 7.1$ degrees elongation from the Sun with J-maps constructed using the \emph{SATPLOT} tool, resulting in prediction lead times of $-26.4 \pm 15.3$ hours. The geometrical models we use assume different CME front shapes (Fixed-$\Phi$, Harmonic Mean, Self-Similar Expansion), and constant CME speed and direction. We find no significant superiority in the predictive capability of any of the three methods. The absolute difference between predicted and observed ICME arrival times is $8.1 \pm 6.3$~hours ($rms$ value of 10.9h). Speeds are consistent to within $284 \pm 288$~\kmsec. Empirical corrections to the predictions enhance their performance for the arrival times to $6.1 \pm 5.0$~hours ($rms$ value of 7.9h), and for the speeds to $53 \pm 50$~\kmsec. These results are important for \emph{Solar Orbiter} and a space weather mission positioned away from the Sun--Earth line. 
\end{abstract}





\section{Introduction}

Storms from the Sun, known as coronal mass ejections (CMEs), are massive, quickly expanding expulsions of plasma threaded by magnetic fields, originating from both quiescent and active regions in the Sun's corona. Over a distance of a few solar radii, they may accelerate up to speeds of 3000 kilometers per second in rare cases, and subsequently propagate through the solar wind away from the Sun. CMEs are able to reach 1 AU in half a day in the most extreme cases, and are the source of the strongest disturbances in the Earth's magnetosphere \citep[e.g.][]{zha07}.

Forecasting their general properties, such as propagation direction and speed close to the Sun and in the interplanetary medium, as well as their arrival time and arrival speed at a given planet or spacecraft, is a major issue in the field of heliophysics. Moreover, the science that underlies our ability to predict CME arrival times and speeds with high precision, which is at the heart of a reliable, real-time space weather forecast, is still not well understood, with average errors of the order of 0.5 to 1 day and several $100$~\kmsec, respectively \citep[e.g.][]{gop01}. Recent advances in analysing multi-point imaging data has improved these classic values for a few CME events by roughly a factor of two \citep{mis13, col13}. While this is clearly of greatest interest for the location of the Earth, predictions and parameters of CMEs impacting Mercury, Venus, Mars, Jupiter and Saturn are also essential for studying their interaction with other planetary environments \citep[e.g.][]{bak13}. 

The \emph{Solar TErrestrial RElations Observatory} \citep[\emph{STEREO}, ][]{kai08}, launched in late 2006, consists of two spacecraft, one ahead (\emph{STEREO-A}) and one behind the Earth (\emph{STEREO-B}) in orbits around the Sun. Each year, each spacecraft separates from the Earth by about 22\degree in heliocentric longitude. Its \emph{SECCHI} instruments seamlessly image CMEs from the Sun to 1 AU and beyond. We are able to extract CME speeds and directions from the images in order to pin down CME evolution and assess CME predictions \citep[e.g.][]{moe09c,woo10,liu10,lug10,  lyn10,sav10,dav11, lie11, moe11, sav12b,rol12, liu13, mis13, col13, dav13}. Even numerical simulations have been employed to enhance our ability to derive the physics of solar wind structures from heliospheric imaging \citep[e.g.][]{lug09a,lug11, xio13b, xio13, rol13}. A concerted campaign to analyse the series of CMEs launched by the Sun on 2010 August 1 has also revealed many details of CME propagation and their 3D evolution for interacting CMEs \citep{liu12,har12,moe12,web13,tem12}.

A deceptively simple but persistent problem in the analysis of CMEs has been to find definite connections between the remote images of CMEs \citep{ill85}, taken by coronagraph or heliospheric imager instruments, and the signatures of CMEs in time series of plasma and magnetic field measurements taken directly (in situ) in the solar wind \citep{bur81}. Current space-based coronagraphs on 3 different spacecraft (\emph{STEREO-A/COR1/COR2, STEREO-B/COR1/COR2, and SOHO/LASCO)} can image a CME in its entirety during its ``birth'' and early propagation phase ($1-15$ solar radii). These instruments provide white-light images of CMEs projected into the plane of the sky, i.e.\ the plane perpendicular to the Sun--spacecraft line.  As the CME propagates further away from the Sun, heliospheric imager (HI) instruments image, in white--light, its integrated density signature around the so-called ``Thomson surface'' \citep{vou06} or ``Thomson plateau'' \citep{how12b} at elongation angles of 4-88\degree from the Sun (for \emph{STEREO/HI}). Because of the wider viewing angle, the interpretation and analysis of HI data is more complex compared to coronagraphs and not yet fully understood \citep[e.g.][]{rou11rev,how12b}.

When a CME hits a spacecraft with in situ instruments onboard that are capable of characterizing the solar wind plasma and magnetic fields, it leads to distinct signatures in the time series of the measured parameters. Often, a shock is followed by a sheath region in front of a magnetic driver, which is either an irregular structure or a large-scale magnetic flux rope \citep[e.g.][]{bur81,bot98,lyn03,lei07,moe09b,ric10,isa13,alh13}. We call the interval including all of these signatures an ``interplanetary CME'' or ``ICME''. Troughout this paper, we use the term ``CMEs'' for events observed in images and ``ICMEs'' for ejecta identified from in situ measurements. However, these measurements of the in situ solar wind give a detailed but otherwise extremely limited and localized view of a CME at the position of the spacecraft near 1 AU. At this distance, a CME has already expanded into an enormous structure, covering up to around 100\degree in heliocentric longitude and several tenths of an AU along the radial direction to the Sun \citep{bur81,bot98,liu05,ric10,woo10,moe12}. Thus, progressing further from the Sun leads to less and less information on the global structure of the CME, which is increasingly hard to interpret, providing a first explanation for the difficulties in linking the datasets.

As a background to our study, it needs to be understood that CMEs are almost never really single entities, but they react to their coronal and heliospheric environments \citep[e.g.][]{tem11, kil12b, rol12}. A possible general paradigm of CME propagation through the solar wind has recently emerged. \cite{liu13} state a picture of Sun--to--Earth propagation of fast CMEs, derived from a joint analysis using stereoscopic heliospheric images, radio and in situ observations of three CME events: fast CMEs impulsively accelerate close to the Sun, followed by a rapid deceleration out to about 0.4~ AU, culminating in an almost constant speed propagation phase \cite[see also, e.g.,][for earlier, similar ideas]{gop01}.

Our paper addresses the problem of seamlessly connecting CMEs from the Sun to 1 AU in two ways. First, we establish a list of suitable CME events from 2008-2012, using observations by coronagraphs, heliospheric imagers and in situ instruments onboard the \emph{STEREO} and Wind spacecraft, and model the imaging observations using single-spacecraft state-of-the-art-``geometrical modeling'' methods. Secondly, we use these connections to enhance our understanding of CME propagation and their prediction, focusing on the improvement of methods to extract CME parameters from heliospheric imager data. Our study  builds on the results of \cite{lug12}, who analyzed CMEs in \emph{STEREO/HI} data that impacted the other \emph{STEREO} spacecraft during the interval 2008--2010. This study established that predictions of speeds and arrival times for slow CMEs are roughly consistent with corresponding in situ parameters, and demonstrated the possibility for succesfully predicting CMEs that propagate behind the limb or on the backside of the Sun as viewed from HI. However, our paper goes further along several, critical avenues: we included modeling with the new Self-Similar Expansion Fitting (SSEF) technique \citep{dav12}, which allows more flexibility for the CME width in the solar equatorial plane than previous models, which were either point-like \citep[Fixed-$\Phi$, ][]{she99,rou08} or extremely wide \citep[Harmonic Mean, ][]{lug09a,lug10}. In addition, we now have \emph{STEREO} data of very fast CMEs ($> 2000$~\kmsec) available, and we mainly discuss Earth-directed events. The \emph{STEREO} separation from Earth increased from 22\degree in early 2008 to 120\degree in mid 2012, thus we include in our study backsided events (from the point of view of a \emph{STEREO} spacecraft), similar to \cite{lug12}.

Several lines of research have recently converged to make this work possible, apart from having available the remote \emph{STEREO} observatories in conjunction with the Wind spacecraft (the latter at the Sun-Earth Lagrangian 1 or L1 point): (1) the existence of software packages (i.e., \emph{SATPLOT} and croissant modeling) to easily extract and fit CMEs in \emph{STEREO/COR2} and HI data, (2) the development of mature models to obtain CME speeds and directions from heliospheric imagers, and (3) the rise of solar cycle 24 which culminated in several very fast coronal mass ejections. All this now gives us the opportunity to study the potentially most geoeffective events with multi--point in situ and imaging data \citep[e.g.][]{liu13,dav13}.
  
The connections established in our paper may be used by other researchers to benchmark  any empirical or numerical propagation model of CMEs. We need to emphasize that our paper focuses on establishing connections between the datasets, which means that we track CMEs as far as possible through the HI data, up to 30--40\degree elongation from the Sun. For this reason the prediction lead times, which are the differences between the last data points used for extracting CME parameters, and the actual in situ arrival times, are relatively short, of the order of one day. This should be kept in mind when discussing our results for the errors in CME speeds and arrival times. Shorter tracks in HI must be used in order to increase the prediction lead time to values suitable for real time forecasting, namely from more than one day for fast CMEs to several days for slow CMEs. Such an analysis was carried out by \cite{moe11} for a case study, and is the aim of future studies that will exploit our full dataset, consisting of 22 CME events. The differences in speeds and arrival times that we derive from comparing the HI data to the in situ data thus show how well a HI system configured as on \emph{STEREO} can perform for CME forecasting.

\section{Methods}

This section illustrates the observations and models we used to extract CME parameters from various imaging and in situ data. We chose the events in our list by the criteria that each CME has (1) clear in situ signatures and (2) a clear track in J-maps from one \emph{STEREO/HI} instrument. We then tracked back the CME in the HI J-map to the coronagraph images to find the corresponding CME observations in the corona. We do not attempt to provide a full list of all CMEs directed towards Earth and \emph{STEREO} during the time between April 2008 and July 2012. However, some of the brightest, fastest and best-studied events in this time-range show up in our list. A well observed, Earth-directed CME that left the Sun on 2012 July 12 and arrived at Earth on 2012 July 14 serves to demonstrate our data pipeline.


\subsection{Croissant modeling}

The so-called ``croissant'' model, also known as ``Graduated Cylindrical Shell'' (GCS) model, is a very well tested tool, available in the \emph{IDL SolarSoft} package, for deriving the initial directions and speeds of CMEs in the fields of view of the \emph{STEREO/COR2} coronagraph, covering 2.5--15.6 $R_{\odot}$ \citep{the06,the09}. At least two viewpoints are needed. Sometimes, the additional viewpoint of the \emph{SOHO/LASCO} C2 and C3 coronagraphs at the near-Earth L1 point is used, to either give the second view or to provide a third. C2 and C3, combined, cover a distance of 1.5--30 $R_{\odot}$ \citep{bru95}. The model has also been recently extended for the fields of view of heliospheric imaging \citep{col13}, but we will use it in the classic way for analyzing coronagraph images only. Essentially, the model approximates the CME shape as a bent tube which is reminiscent of a croissant, with the mass distributed uniformly on its surface. This shape was chosen to be consistent with a geometry of the CME internal magnetic field as a bent, cylindrical magnetic flux rope. The application of the model consists of manually adjusting parameters that define the CME shape and direction to best fit images of the CME provided by coronagraphs. The major advantage of using this model is that it provides reliable results for the initial 3D propagation direction and initial speed of a CME, free of projection effects. The downside is that not every CME can be approximated by the shape of a croissant, and often CMEs occur within a few hours of each other, making it difficult to distinguish the different physical parts of each CME. Nevertheless, \cite{vou13} found that, close to the Sun, at least $40$\% of CMEs possessed a flux-rope like structure. We refer the reader to an extensive discussion on the morphology of CMEs in coronagraphs that is provided by those authors.

\begin{figure}[t]
\epsscale{1.2}
\plotone{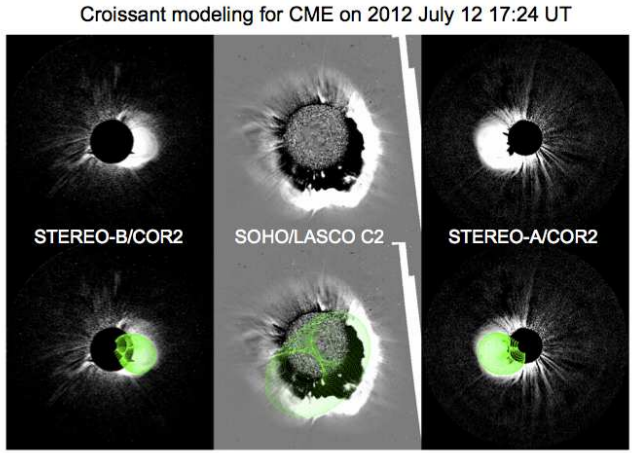}
\caption{Croissant modeling of the 2012 July 12 CME on 17:24 UT. From left to right: \emph{STEREO-B/COR2}, \emph{SOHO-LASCO C2}, and \emph{STEREO-A/COR2}. The top row shows the white-light image, the bottom row a green grid of the tube-like croissant model overlaid. The separation of \STB to Earth was 115\degree in heliocentric longitude, and \STA was 120\degree ahead of Earth.}\label{croissant_july}
\end{figure}

As an example, the top row in Figure~\ref{croissant_july} shows images of the 2012 July 12 17:24 UT CME, and the bottom row a green grid of the croissant model overlaid on the same images. It can be seen that the croissant represents the shape of the  CME well in all 3 viewpoints. It becomes clear that this CME is roughly directed towards the Earth, appearing as a full halo in \emph{SOHO/LASCO}, and being directed to solar west from \emph{STEREO-B}, and to solar east from \emph{STEREO-A}. To get the CME initial speed $V_{\mathrm{init}}$, we linearly fit height-time measurements for the croissant's apex over a few data points \citep[e.g.][]{col13}. Thus, $V_{\mathrm{init}}$ is an average speed between 2.5--15.6 $R_{\odot}$. The initial CME direction is similarly obtained by adjusting the wire-grid model to the images, and averages over a few data points define the initial CME direction in heliocentric longitude ($\Phi_{\mathrm{init;Earth}}$) in the solar equatorial plane, with the Earth as a reference point at~0\degree longitude.

Common errors associated with croissant modeling are  $\pm$10\degree in heliocentric longitude for the CME propagation direction, and $\pm$10\% of the average CME speed  \citep[e.g.][]{the09,tem12}. In Table~1, all results derived with the croissant model are summarized in columns 2--4, including the time $t_{\mathrm{COR2}}$ of the first observation in \emph{STEREO-A/COR2} images, the initial direction with respect to Earth ($\Phi_{\mathrm{init;Earth}}$) and the initial speed ($V_{\mathrm{init}}$). For events 16 to 24 in Table~1, we have used 3 viewpoints (\emph{STEREO-A/B} and \emph{SOHO}), and for the other events the 2 viewpoints as provided by \emph{STEREO-A/B}. For Earth-directed CMEs, which appear symmetric in \emph{STEREO-A/B}, an additional viewpoint provides more reliable results \citep{vou11}.

\subsection{Geometrical modeling}

\begin{figure}
\epsscale{1.25}
\plotone{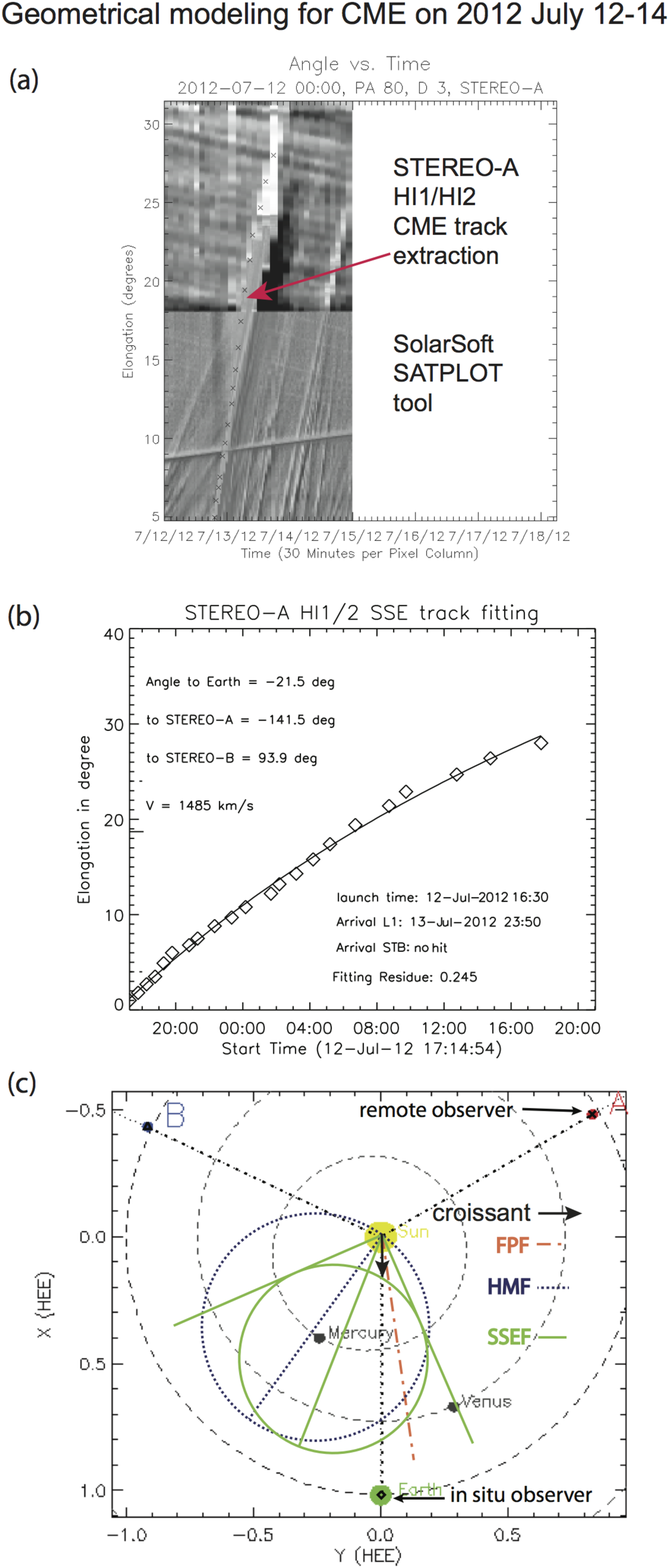}
\caption{Geometrical modeling of the 2012 July 12-14 CME. (a) The density track of the CME visible in a J-map from \emph{STEREO-A}, with extracted time and elongation data points (``x'' symbols) of the CME front, using the \emph{SATPLOT} software tool available in IDL \emph{SolarSoft}. (b) Fit of the extracted CME track with the Self-Similar Expansion Fitting (SSEF) model. Some results are indicated on the plot. Note that the Fixed-$\Phi$ Fitting (FPF) and Harmonic Mean Fitting (HMF) models (not shown) also reproduce well the observed time-elongation track. (c) The resulting geometry of the event, with propagation directions derived from FPF (dot-dashed red line emanating from the Sun), SSEF (solid green line) and HMF (dotted blue line) indicated. The HMF circle (180\degree full width) is dotted blue, and the SSEF circle (90\degree full width) is solid green.  Also indicated is the direction from the croissant modeling by a black arrow. }\label{summary_july}
\end{figure}

Figure~\ref{summary_july} is a summary of the application and the results of the geometrical modeling techniques. Figure~\ref{summary_july}a shows an elongation versus time plot or J-map  \citep[e.g.][]{she99,rou08,dav09,liu10b} for \STA on 2012 July 12--15. In this plot, a stripe of a combined \emph{STEREO/COR2/HI1/HI2}  image is extracted along a given position angle (PA), indicating the solar latitude which is presented. The stripes are then put together vertically one after another to form a J-map (with a 30 minute time interval).  The PA is measured from solar north (0\degreee) to east (90\degreee), south (180\degreee) and west (270\degreee). Consequently, when using \emph{STEREO-A(B)} images, a PA of around 90\degree (270\degreee) should be chosen to obtain measurements in the solar equatorial plane, which is also close to the ecliptic plane where the in situ observations are made. 

In reality, CMEs often propagate away from the solar equatorial plane \citep[e.g.][]{kil09c}, and their signature is too faint around PA=90\degreee. For such CME events we choose a PA between 77\degree and 95\degreee, depending at which PA we were able to capture the CME front clearly. For the J-map in Figure~\ref{summary_july}a, PA=80\degreee, using data taken within $D= \pm 3$\degree of this PA. This is also the reason that we will not further distinguish between measurements in the solar equatorial plane (valid for the  croissant results) and ecliptic plane (valid for in situ results), because the HI data connecting the solar to in situ measurements are not exactly taken in either of these planes. This does not affect our conclusions because the resulting systematic error in heliocentric longitude of a few degrees is much lower than the differences between CME directions given by the different HI models, of the order of $> 10$\degreee.

Figure~\ref{summary_july}a was created with the \emph{SATPLOT} software package, freely available in \emph{IDL SolarSoft}. With this tool, a user can produce J-maps and movies from data from both \emph{STEREO/HI} instruments, and interactively extract time-elongation $\epsilon(t)$ functions of dense solar wind structures by clicking in the J-map (see the online manual for more information on the \emph{SATPLOT} package). In our study, these are measurements of the CME leading edge, which is the front of a grey or white trace in the J-maps indicating enhanced solar wind densities. This signal arises for a HI observer due to Thomson scattering of photospheric light off solar wind electrons, over a range of distances along the line-of-sight \citep{vou06,how12b}. 

Figure~\ref{summary_july}b shows the result of fitting the extracted $\epsilon(t)$ track using the Self-Similar Expansion Fitting or SSEF model \citep{dav12,moe13}. This provides, for the time and elongation range of the HI observations, an average CME speed and its direction, as well as its arrival time at Earth and the other \emph{STEREO} spacecraft. This fitting procedure works autonomously and takes only a few seconds. We also use two other models, the Fixed-$\Phi$ Fitting \citep[FPF,][]{she99,rou08} and Harmonic Mean Fitting \citep[HMF, ][]{lug10b,moe11} techniques, which are both available in \emph{SATPLOT}. It is important to emphasize that these methods are all used for a \emph{single-spacecraft HI observer}. Essentially, all of these models are based on purely geometrical considerations, and use the concept that deceptive acceleration and deceleration arises in $\epsilon(t)$, depending on the speed and direction of the CME with respect to the HI observer \citep[see, for example, extensive descriptions in][]{lug10b,moe11,dav12,lug13}. We appropriately call them ``geometrical models''. \emph{All methods share the same assumptions of constant CME speed and direction, but differ on the description of the global shape of the CME front.}

Figure~\ref{summary_july}c shows the resulting CME directions for the 2012 July 12-14 CME event in the interplanetary medium, and also serves as an illustration of the different geometries. The FPF model assumes a point-like CME without any extension in heliocentric longitude (the dot-dashed red line). The HMF model is shown as a very wide blue circle, with $\lambda=90$\degreee, where $\lambda$ is the CME half width \citep[e.g.][]{dav12,moe13}.  The SSEF model can be seen as a generalization of the FPF and HMF models, because $\lambda$ can vary anywhere between 0 and 90\degreee, which correspond to FPF and HMF, respectively. In our analysis, every result for the SSEF model uses a half width of $\lambda=45$\degreee, as this forms the average value between the other two models. It is illustrated by the green circle in Figure~\ref{summary_july}c. 

For the longitudinally extended HMF and SSEF models, the speed to be compared to the in situ speed is not the apex speed from the geometrical model ($V_{\mathrm{IP}}$), but a different speed calculated for the point on the CME model front that is directed towards the in situ observatory. We call it $V_{\mathrm{IPo}}$, for interplanetary (IP) speed toward the observer (o). In Figure~\ref{summary_july}c, this point is the intersection of the green SSEF circle with the Sun-Earth line, close to Earth. For the FPF model a point-like CME shape is assumed, and it is not possible to differentiate between the two speeds. The speeds $V_{\mathrm{IPo}}$ can be calculated by knowing the apex speed and the difference in heliocentric longitude between the CME apex in the model and the position of the in situ spacecraft. The corresponding formulas are derived and summarized by \cite{moe13}. 

There is an error associated with manual measurements of CME fronts in J-maps, which is on the order of $\pm$0.5\degree in elongation angle $\epsilon(t)$ \citep{wil09,liu10,moe11}. Depending on the length of the $\epsilon(t)$ track, this error will propagate into the results of the geometrical models. In general, we tracked each CME to about 35\degree elongation from the Sun, because this is a range for which the fitting methods are known to give relatively well constrained results \citep{wil09, lug11, moe11}.Because the tracks need to cover a large range in elongation for the fitting techniques to work, the methods cannot be applied to coronagraph measurements.

Concerning interplanetary CME directions, $\pm$0.5\degree in $\epsilon(t)$ translates to an error of $<\pm 5$\degree  for CMEs tracked further than 30\degree elongation from the Sun \citep{wil09,lug11,moe11}. Thus we choose $\pm 5$\degree as the error bar on all plots including directions from geometrical modeling. For the interplanetary speeds $V_{\mathrm{IP}}$, we experimented with extracting several tracks of a $\approx 1000$~\kmsecc CME in J-maps created with \emph{SATPLOT}, and found a conservatively chosen error of $\pm 10$\% in $V_{\mathrm{IP}}$, that we use on all plots that show interplanetary speeds. A similar range was found by \cite{lug11}  through numerical testing. Our experiment showed systematic errors in the arrival times of a maximum of $\pm 10$\% of the total CME transit time.

Table~1 presents a summary of results from geometrical modeling. We state, for simplicity, the results from SSEF only (with $\lambda= 45$\degreee). The speeds and directions from SSEF are in between those from the extreme models FPF and HMF, thus the SSEF results form a good average of these parameters. Columns 5 to 10 of this table show the interplanetary CME direction (heliocentric longitude) $\Phi_{\mathrm{IP;Earth}}$ with respect to Earth and with respect to the HI observer, $\Phi_{\mathrm{IP;HI}}$, as well as the speed of the model apex, $V_{\mathrm{IP}}$, and the speed of the front in the direction of the in situ observatory, $V_{\mathrm{IPo}}$. Finally, $t_a$ is the predicted arrival time of the CME leading edge at the 
spacecraft stated in the last column.

\subsection{In situ solar wind data}

\begin{figure*}[th]
\epsscale{1}
\plotone{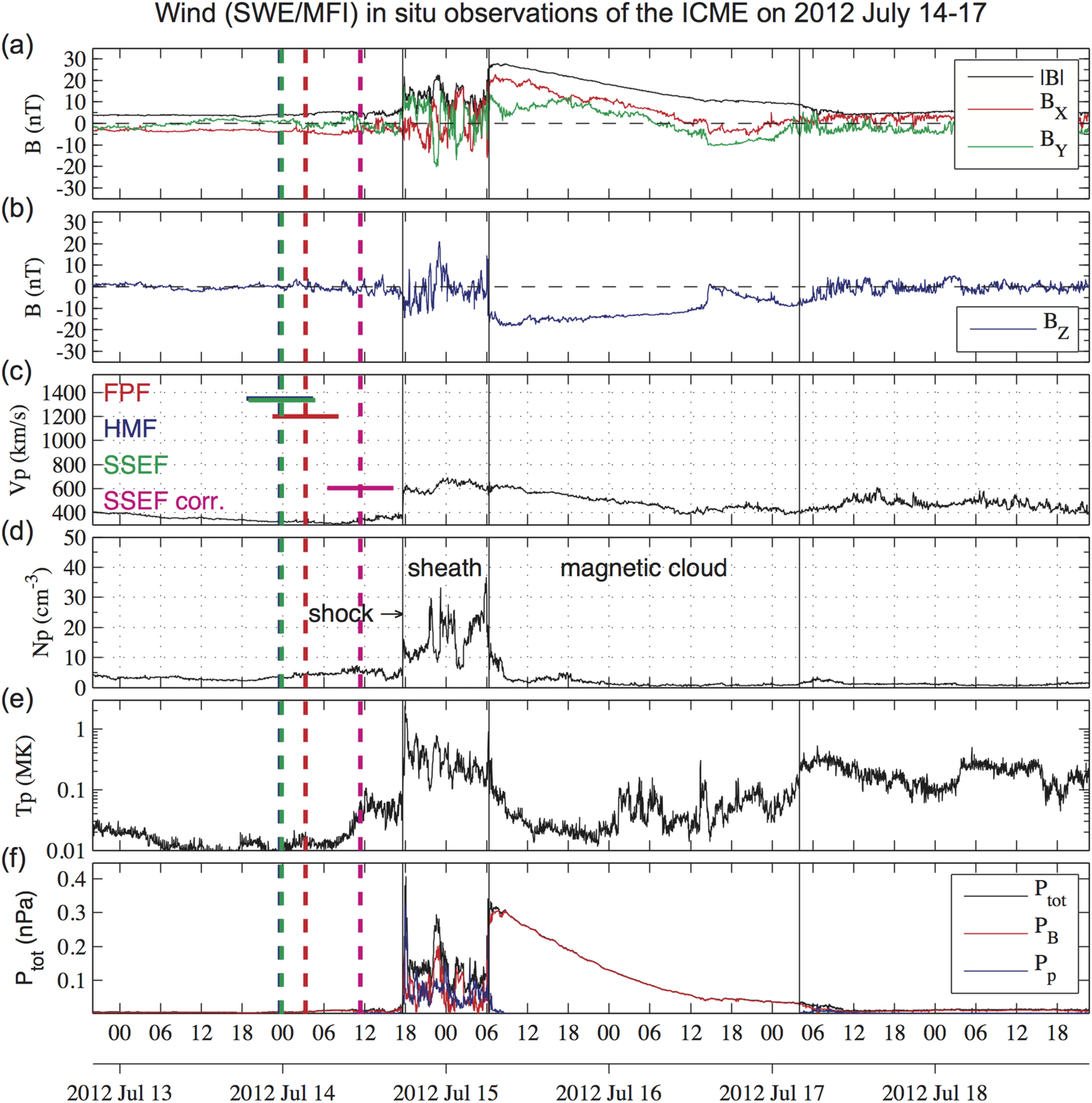}
\caption{In situ solar wind observations from the near-Earth Wind spacecraft, at the Sun-Earth L1 point, between 2012 July 13--18. From top to bottom: (a) Total magnetic field strength and $B_X$ and $B_Y$ components (\emph{in Geocentric Solar Ecliptic} or GSE coordinates). (b) Magnetic field component $B_Z$ in GSE. (c) Proton bulk speed. (d) Proton number density. (e) Proton temperature. (f) Magnetic ($P_B$), plasma ($P_{\mathrm{p}}$) and total pressure ($P_{\mathrm{tot}}$). Predicted arrival times from the FPF (red), HMF (blue) and SSEF (green) models are indicated as vertical dashed lines. The speeds predicted by the same models are shown in panel (c) as horizontal bars, in similar colors, for direct comparison to the in situ proton bulk speed. The width of the horizontal bars corresponds to the estimated error in arrival time resulting from the manual selection of points ($\pm 10$\% of the total CME transit time). The corrected arrival time (in version 1, see text) and the corrected speed (``SSEF corr.'') from section \ref{connectingsection} are shown as pink vertical and horizontal lines, respectively.
}\label{insitu_july}
\end{figure*}

Figure~\ref{insitu_july} presents an overview of the near-Earth (L1) in situ solar wind data of proton bulk parameters and magnetic field components from 2012 July 13--18.  We show data from the \emph{ Solar Wind Experiment} \citep[\emph{SWE}, ][]{ogi95} and the \emph{Magnetic Field Investigation} \citep[\emph{MFI},][]{lep95} on the \emph{Wind} spacecraft, at a 1-minute time resolution. For 5 of the 24 in situ ICME arrivals in our study we use magnetic fields and proton data from the \emph{IMPACT} \citep{luh08} and \emph{PLASTIC} \citep{gal08} instruments on the \emph{STEREO-B} spacecraft.

In Figure~\ref{insitu_july} we can see the signatures of an ICME in the near-Earth solar wind. A clear shock is seen on 2012 July 14 1738 UT, signaled by sudden jumps in magnetic field, speed, density and temperature, delimited by the first solid vertical line on the left. Behind the shock follows the sheath region of high density and high temperature solar wind, and variable magnetic field. Around 2012 July 15 0600 UT, at the second solid vertical line, this region ends and the interval of a magnetic cloud \citep{bur81} begins, which extends about 48 hours up to early July 17, where a third vertical line signals the end of the cloud. This region is an example of a clean magnetic structure passing the spacecraft, and is characterized by strong magnetic field strength, a smooth rotation of the magnetic field vector, which is shown in \emph{Geocentric Solar Ecliptic} (\emph{GSE}) coordinates, and low proton temperature. Such observations are usually interpreted as a magnetic flux rope, extending tube-like from the Sun with a helical magnetic field geometry \citep[e.g.][]{alh13, jan13}. We do not discuss magnetic cloud geometry further in this paper, but, for completeness, in this case the field rotates from solar east ($B_Y > 0$) to south of the ecliptic ($B_Z < 0$) to solar west ($B_Y < 0$). This cloud is of east-south-west (ESW) type \citep{bot98, mul98}, and its axis is consequently roughly normal to the ecliptic plane pointing southward. Also shown are the results for arrival times and speeds by the geometrical modeling methods (``SSEF corr.'' will be explained in Section 3.2). The fact that they are accurate to within a few hours of the observed arrival time at the location of Earth establishes the connection from the remote HI data to the in situ data near 1 AU.

In Table 2, columns 5--9, we present in situ results for every ICME event in our dataset, derived from similar plots. These parameters are: (column 5) the arrival time $t_{\mathrm{insitu}}$ of the shock, or, if no shock is present, the arrival time  of a significant jump in proton density, (6) the average proton bulk speed $V_{\mathrm{sheath}}$ inside the sheath region, and its standard deviation, (7) the average proton density $N_{\mathrm{sheath}}$ inside the sheath, and its standard deviation, (8) the maximum magnetic field $B_{{\mathrm{max}}}$ in the full ICME interval (consisting of shock, sheath, and any magnetic structure), (9) the minimum value of the southward component of the magnetic field (min $B_Z$) in the full ICME interval. Some of these parameters act as an independent check on the validity of the results of the HI models, and highlight differences between the datasets when connecting CME observations from the Sun to 1 AU.

For this comparison, most relevant are the average proton bulk speed inside the ICME sheath region and the arrival time of the shock. The reason that we choose these parameters is that previous research has shown that the density, which is mapped by the HI instrument, and which shows up as a high density ``track'' in J-maps, corresponds to the region of enhanced plasma density in the ICME sheath \citep[e.g.][]{rou09,moe09c,moe10,lyn10,liu11,how12}. Thus, tracking the leading edge of a CME in a J-map essentially means tracking the front of the sheath region, which is the location of the shock. Hence, the predicted arrival time from geometrical modeling corresponds to the  shock arrival time observed in situ, and $V_{\mathrm{IPo}}$ corresponds to $V_{\mathrm{sheath}}$.

These in situ parameters are excellently suitable to act as target values for understanding the performance and limitations of the geometrical models, because they are very well defined, and have very small intrinsic errors. The arrival time of the shock can be pinpointed usually to within a few minutes accuracy, which is negligible when compared to the several hour error in arrival time prediction. For very slow ($\approx 400~$\kmsec) CMEs, where no shock is usually present, the in situ arrival time can be defined by the start of a high density region in front of the magnetic structure, to an uncertainty of $\pm 5$ hours. This definition was necessary only for the events with numbers 4 and 10 in our list. The speed $V_{\mathrm{sheath}}$ (column 6 in Table 2) has an average  standard deviation of $\pm 23$ \kmsec. This is only 5\% of the average sheath speed of 485 \kmsec, so it is also a well constrained quantity.

\section{Results}

In this section, we discuss the results from an analysis of all the parameters that we described in the previous section for the full set of 22 different CME events. For the 2008 June 1--7 CME, there are two tracks in HI, one for the leading and one for the trailing edge, enveloping a magnetic flux rope \citep{moe09c, lyn10}, thus we have two HI/in situ comparisons for a single CME. Additionally, for the CME on 2010 August 1--4, in situ data are available from the \WIN and \STB spacecraft, which observed different parts of the same shock front associated with the fastest and largest of several CMEs launched on the Sun on 2010 August 1 \citep{liu12,har12, moe12,web13}. In total, we end up with 24 different comparisons of imaging to in situ data. To track CMEs in HI data, we use \emph{STEREO-A} for 21 events and \emph{STEREO-B}  for 1 event because of the slightly higher image quality for \emph{STEREO-A} and some data gaps for \emph{STEREO-B}. The imaging parameters are summarized in Table~1, and the in situ parameters are given in Table~2.

\subsection{Connecting coronagraph to HI data}

In this section, we discuss the comparison between CME parameters derived from the different imaging methods of croissant and geometrical modeling (Table~1). Essentially, we look for consistency between the CME speed and direction in the outer corona ($< 15~R_{\odot}$ or 0.07~AU) and in the interplanetary medium (roughly up to 0.85 AU, see section 3.2.2 for details on the distance range). The initial directions and speeds of the CMEs are very well defined in the corona, as they are constrained by multi-view point imaging; this serves as a point of reference.

\subsubsection{CME directions}

\begin{figure*}
\epsscale{1.0}
\plotone{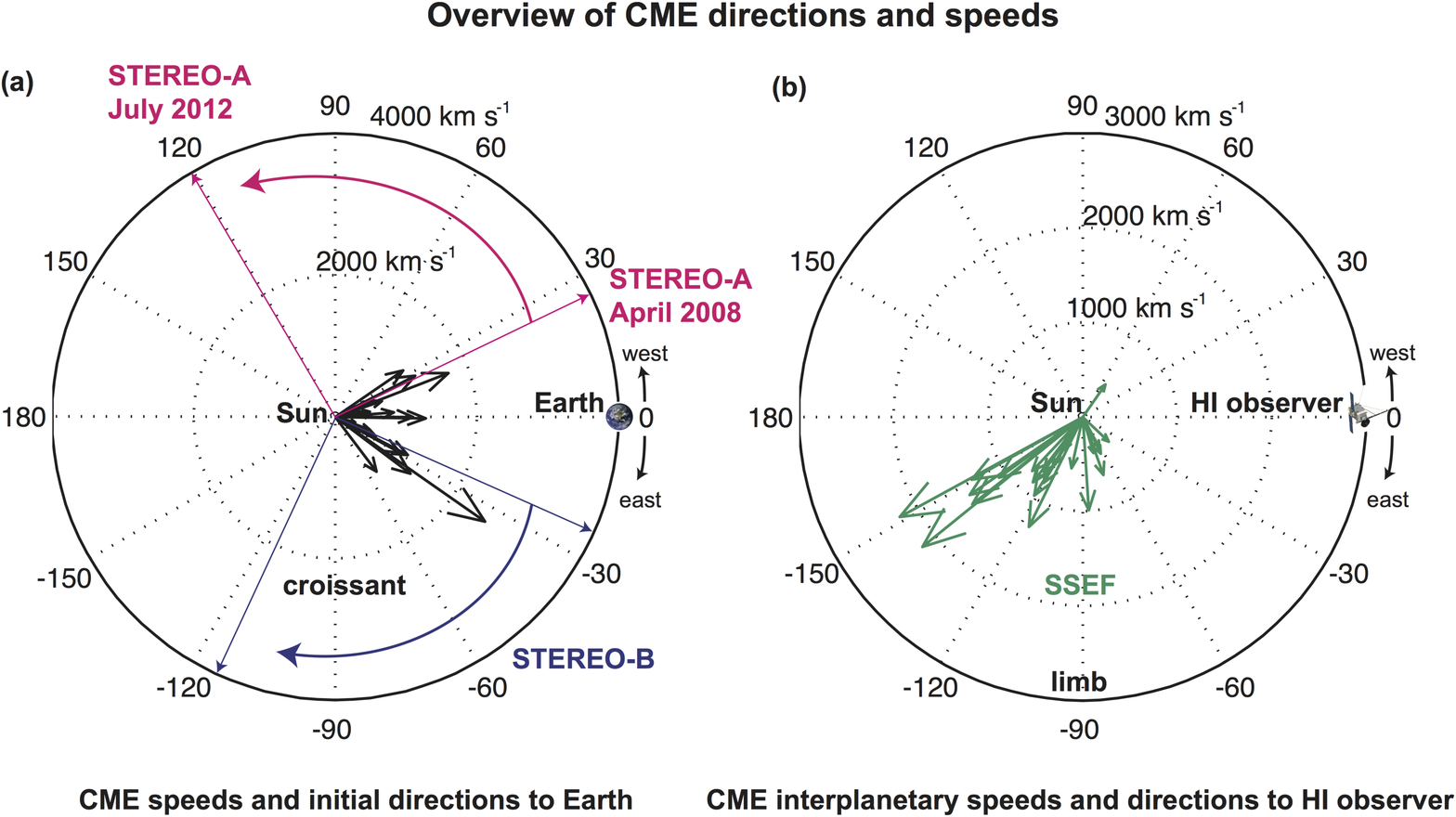}
\caption{(a) CME initial directions ($\Phi_{\mathrm{init}}$) in the solar equatorial plane, with respect to Earth (at 0\degree longitude). The length of each arrow indicates the CME initial speed ($V_{\mathrm{init}}$). Both values are derived from croissant modeling, for a distance range $2.5-15~R_{\odot}$. (b) CME interplanetary directions ($\Phi_{\mathrm{IP}}$) from the SSEF model, in planes close to the solar equatorial plane, visualized with respect to the HI observer, which is \emph{STEREO-A} for all but one event, explaining the bias towards solar east. The arrow length indicates the interplanetary CME speed ($V_{\mathrm{IP}}$) derived from SSEF modeling. It can be seen that for most of the fast CMEs the SSEF models yields backsided directions.}\label{direction_all}
\end{figure*}

Figure \ref{direction_all}a shows the initial directions $\Phi_{\mathrm{init;Earth}}$  in the solar equatorial plane of all CMEs in our dataset, with the length of the arrow refering to the initial speed, both derived from the croissant model. In the time between the first event in 2008 April and the last event in 2012 July, \STA separated itself from Earth from 25\degree to 120\degreee. It can be seen that for our specific set of events, the fastest CMEs tended to be launched about 30\degree east and west of Earth, and only a few events were initially directed centrally toward Earth.

Figure \ref{direction_all}b illustrates the interplanetary directions from SSEF modeling ($\Phi_{\mathrm{IP;HI}}$), relative to the HI observer, which is at the fixed longitude of 0\degreee. The length of the arrow corresponds to the interplanetary speed of the SSEF circle's apex ($V_{\mathrm{IP}}$). This figure demonstrates the important point that the fastest CMEs in our dataset were all backsided from the vantage point of the HI observer, and that the slow events were mainly frontsided. This is simply a consequence of the weak solar cycle 24, which resulted in CMEs which had well over 1000 \kmsecc only after the end of 2011. By this time, the two \emph{STEREO} spacecraft had already reached a mutual separation greater than 180\degree.

\begin{figure*}
\epsscale{1.2}
\plotone{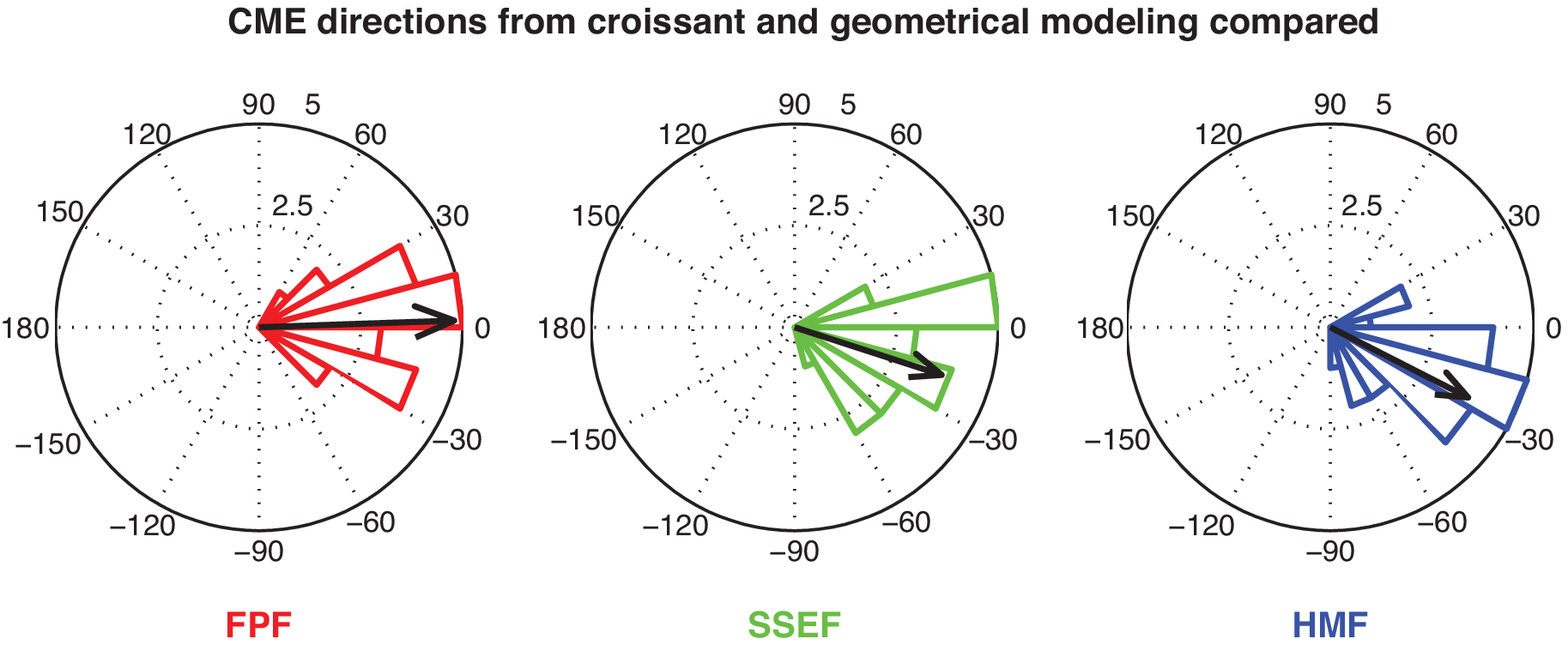}
\caption{A comparison of CME propagation directions in heliocentric longitude, from croissant modeling ($\Phi_{\mathrm{init;Earth}}$) and geometrical modeling ($\Phi_{\mathrm{IP;Earth}}$) for, from left to right, FPF, SSEF, and HMF. For each CME, the difference ($\Delta \Phi= \Phi_{\mathrm{IP;Earth}}- \Phi_{\mathrm{init;Earth}}$) is calculated, and the histogram of resulting angular differences is presented in bins of 15\degree width. The direction of 0\degree thus represents the croissant direction ($\Phi_{\mathrm{init;Earth}}$) in each panel. The black arrows show the mean of the direction difference. As the assumed width of the CME in the models rises from left to right, the  difference in direction between the croissant and geometrical model increases, in the direction away from the imaging observer. }\label{direction_polar}
\end{figure*}

Figure \ref{direction_polar} compares the CME directions from croissant modeling, $\Phi_{\mathrm{init;Earth}}$, to those from each geometrical model, $\Phi_{\mathrm{IP; Earth}}$. Each panel presents an angular histogram of  $\Delta \Phi = \Phi_{\mathrm{IP; Earth}}-\Phi_{\mathrm{init;Earth}}$ with bins of 15\degree width;
$\Phi_{\mathrm{init;Earth}}$ is at 0\degree longitude.
The black arrow is the average $<\Delta \Phi>$. There is considerable scatter; as an example, 90\% of interplanetary directions given by FPF are within $\pm 35$\degree of the croissant direction.

The black arrow on each histogram shows that, on average, the croissant direction is consistent with FPF modeling to within a few degrees, whereas for the SSEF and HMF models, there is a bias in direction which increases with larger CME width. It results in an average difference in 30\degree longitude between croissant and HMF directions. Because almost all events are observed by \emph{STEREO-A}, which is positioned to the west of Earth and looks towards solar east, the resulting direction bias is towards solar east or angles $<0$\degreee. This means that the SSEF and HMF methods places the CME apex further away from the HI observer. \cite{lug13}  showed theoretically that this behaviour ultimately results from the constant speed assumption of the models, because the real deceleration of a fast CME in the interplanetary medium is interpreted by the models as geometrical deceleration, which means a change in direction as compared to a CME which is not decelerating. \cite{lug13} argued that the FPF model is in this way superior to the others, because the error resulting from neglecting deceleration is cancelled by neglecting the CME width.

\begin{figure}
\epsscale{1.2}
\plotone{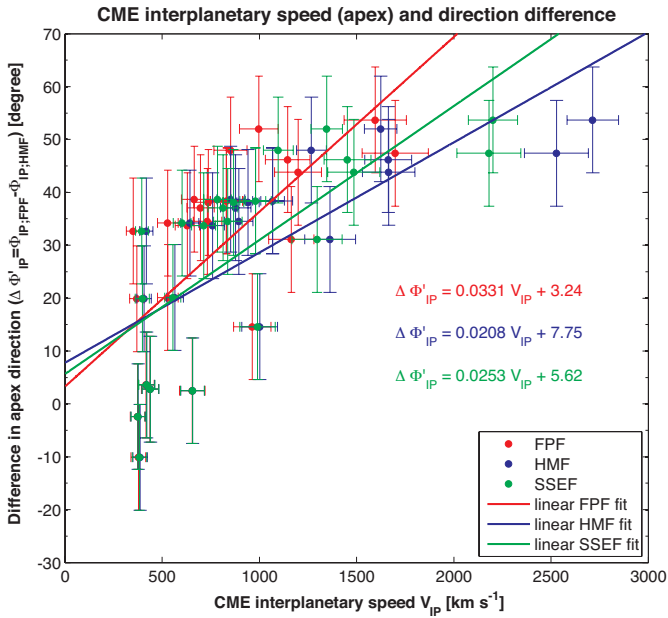}
\caption{CME interplanetary speed $V_{\mathrm{IP}}$ versus difference in  heliocentric longitude of CME propagation directions between the FPF and HMF methods: $\Delta \Phi'_{\mathrm{IP}} = \Phi_{\mathrm{IP;FPF}}-\Phi_{\mathrm{IP;HMF}}$. For each event, $\Delta \Phi'_{\mathrm{IP}}$ is plotted against $V_{\mathrm{IP}}$ of all three methods (FPF, HMF, SSEF), and separate linear fits are performed for each method, with the resulting relationships quoted on the figure.   } \label{direction_difference} 
\end{figure}

We confirm this relationship from the observations in our dataset in Figure~\ref{direction_difference}, where we plot the interplanetary speed $V_{\mathrm{IP}}$ against the difference in direction from the two extreme models FPF and HMF: $\Delta \Phi'_{\mathrm{IP}}=\Phi_{\mathrm{IP; FPF}}-\Phi_{\mathrm{IP;HMF}}$. Higher interplanetary speeds are clearly correlated with larger differences in direction. We also quote linear relationships on the plot from which the resulting FPF--HMF direction difference $\Delta \Phi'_{\mathrm{IP}}$ can be estimated, if $V_{\mathrm{IP}}$ is already known from one of the geometrical models.

We can take away from this section that connecting CME directions from the corona to the interplanetary medium works to within 30\degree in heliocentric longitude. However, the models that feature an extended CME width (SSEF and HMF) and which are at first glance more mature than the point-like FPF model, produce systematically biased directions for fast CMEs. 

\subsubsection{CME speeds and launch times}

\begin{figure}
\epsscale{1.2}
\plotone{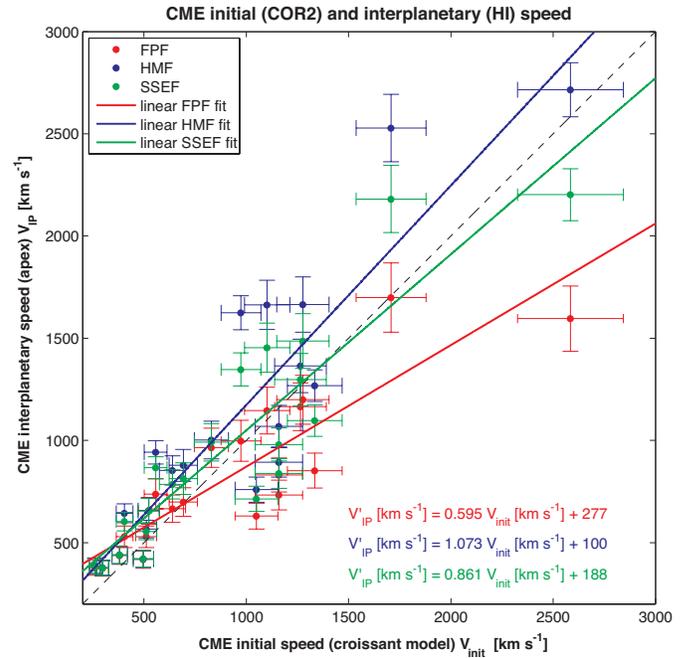}
\caption{Comparison between CME initial speeds $V_{\mathrm{init}}$, as derived from croissant modeling, and interplanetary propagation speeds. For each CME and its initial speed, the corresponding speed derived from the three geometrical models (FPF, HMF, SSEF) is plotted. The results of linear fits for all three models (red solid line for FPF, blue for HMF, and green for SSEF) are quoted with the corresponding colors.}\label{speeds_cor_hi}
\end{figure}

Figure \ref{speeds_cor_hi} plots the relationship between the initial ($V_{\mathrm{init}}$) and the interplanetary ($V_{\mathrm{IP}}$) CME speed of the model apex. For each $V_{\mathrm{init}}$, the three different speeds by the geometrical models are shown. The black dashed line indicates where $V_{\mathrm{init}}=V_{\mathrm{IP}}$. We linearly fitted the data for each geometrical model separately. These linear fits work well for this relationship, and they allow one speed to be estimated if the other is known. They result in:

\begin{gather}
V'_{\mathrm{IP}}=0.595 \; V_{\mathrm{init}}+277, ~  ~ ~ ~ (\text{FPF model})\\
V'_{\mathrm{IP}}=1.073 \; V_{\mathrm{init}}+100, ~  ~ ~ ~ (\text{HMF model})\\
V'_{\mathrm{IP}}=0.861 \; V_{\mathrm{init}}+188, ~  ~ ~ ~ (\text{SSEF model})
\end{gather}
with both speeds in \kmsec.

The FPF method yields a deceleration between the corona and the interplanetary medium (IP), with the slope of the linear fit indicating $V_{\mathrm{IP}} \propto 0.6~V_{\mathrm{init}}$. When using SSEF, the deceleration is less ($V_{\mathrm{IP}} \propto 0.86~V_{\mathrm{init}}$), and for HMF, the speed remains approximately constant, $V_{\mathrm{IP}} \propto 1.07~V_{\mathrm{init}}$. Here, again, our physical interpretation depends highly on which models we use, and thus these relationships are most useful for understanding the biases inherent in the different  techniques.  Nevertheless, taking these results at face value and  knowing from theoretical considerations \citep{lug13} that FPF should give more accurate results for fast CMEs, a deceleration of CMEs from the corona to the IP medium seems to be the most realistic conclusion.

Furthermore, concerning another quantitative link between the \emph{COR2} and HI data sets, we checked whether the launch time $t_0$ resulting from the geometrical models is consistent with the time $t_{COR2}$ (see Table~1) of the first image of the CME in \emph{COR2}. The launch time is defined as the time when the fitting function $\epsilon(t_0)=0$ \citep{moe11}. The average difference $<|t_{COR2}-t_{0}|>=1.7$~hours, and for $92\%$ of all events it is $< 3~$hours. This means that the launch time is an excellent proxy for linking a CME track, which was extracted and from HI J-maps and modeled, to its coronal counterpart, provided that only one CME is launched during this time window.

\subsection{Connecting HI to in situ data} \label{connectingsection}

In this section we discuss the link between the ICME in situ data and the speeds and arrival times derived from the HI observations. The CME direction cannot be compared to in situ because there is no simple way to derive the principal direction of an ICME from in situ data. This might be possible with deeper investigations on orientations of the shocks and flux ropes inside the ICMEs \citep[e.g.][]{liu10b,moe12,isa13}, but this is beyond the scope of the current study. 

\subsubsection{CME and ICME speeds}

We first focus on a comparison of speeds. For the heliospheric imager data, these are the speeds toward the in situ observer ($V_{\mathrm{IPo}}$) derived from the three geometrical models (FPF, HMF, SSEF). For the in situ ICME data, these are the average proton bulk speeds in the ICME sheath regions ($V_{\mathrm{sheath}}$).

\begin{figure*}
\epsscale{1.1}
\plotone{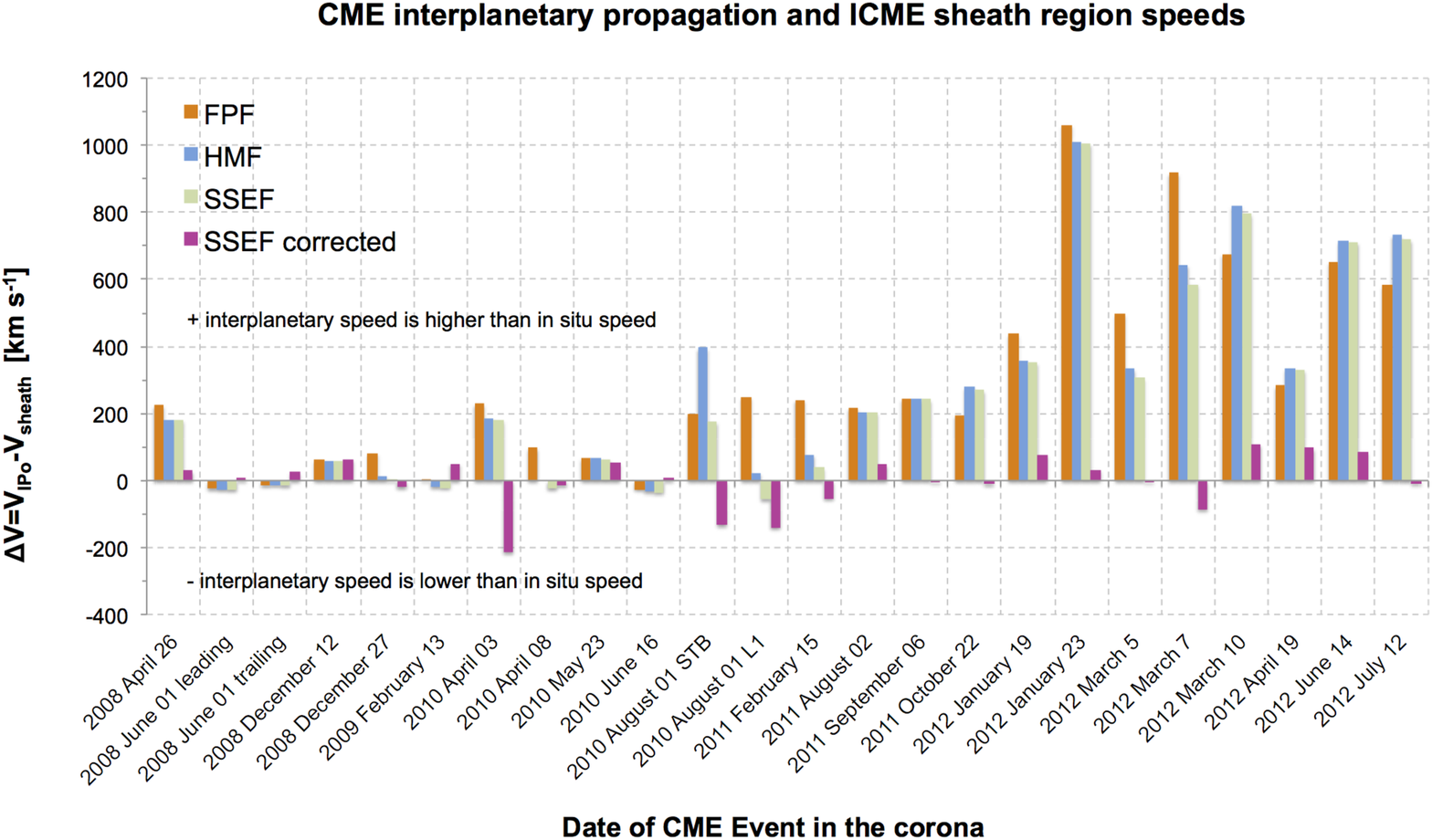}
\caption{Difference between the CME interplanetary speed in the direction of the in situ observatory ($V_{\mathrm{IPo}}$) and the ICME sheath speed ($V_{\mathrm{sheath}}$), for each model (FPF red, HMF blue, SSEF green), and for each CME event. The pink bars show the difference to $V_{\mathrm{sheath}}$ using the corrected speed $V'_{{\mathrm{sheath}}}$ for SSEF modeling. The corrected speed is calculated by using the linear fit for SSEF as indicated on Figure~\ref{speed_scatter}. } \label{speed_bar}
\end{figure*}

Figure \ref{speed_bar} shows the difference $\Delta V =V_{\mathrm{IPo}}-V_{\mathrm{sheath}}$ for each event included in our study. Note that the speed $V_{\mathrm{IPo}}$ already includes a correction to account for which part of the CME front impacts the spacecraft. It is clearly seen that for most slow CMEs in the sample, until late 2011, there is a relatively good consistency. For FPF, in 42 \% of events the absolute difference $|\Delta V|< 200$~\kmsecc (SSEF: 54 \%; HMF: 50\%). This means that for about half of the events there is a reasonable match between the imaging and in situ speeds.

However, most IP speeds exceed the in situ ones. For FPF, 87 \% of events are faster in the IP medium than when observed in situ (SSEF: 75 \%, HMF 83 \%). This is most evident for the fast CMEs in our sample, those with initial speeds $>1000$~\kmsec, for which the predicted in situ speeds are strongly overestimated \citep[consistent with][]{lug13}, on the order of 500 to 1000 \kmsec.  These are the events on the right side of Figure \ref{speed_bar}. The geometrical models yield an average interplanetary propagation speed of a CME. Especially for very fast CMEs, these speeds will not match the speeds measured in situ at 1 AU because of the CME deceleration that occurs up to 1 AU \citep[e.g.][]{gop01, vrs13}. In essence, the assumption of constant speed in the geometrical models is at the root of this discrepancy. 

Table~3 summarizes the $|\Delta V|$ obtained with the different methods. On average, $<|\Delta V|>$ is 303~\kmsec~(FPF), 267~\kmsec~(SSEF) and 282~\kmsec~(HMF).  Taking the average over these values gives 284~\kmsec. We can also see that the differences between the geometrical models are only about 40~\kmsec, which makes it impossible to state which one performs best. In Table~3 we also state the results of a comparison $|\Delta V| =|V_{\mathrm{IP}}-V_{\mathrm{sheath}}|$, which means that we just use the CME model apex speed without a correction for the position of the in situ observer. This results in an average difference over all models of 431~\kmsec. Consequently, correcting for apex or flank encounters strongly enhances the consistency with in situ data, for the present set of events on average by $\approx 150$~\kmsec. However, this is likely caused by the effect that a flank speed is invariably lower than the apex speed, simulating to some degree the effect of real CME deceleration. But even when using $V_{\mathrm{IPo}}$, the average difference to the in situ speeds is still too large, almost 300~\kmsec. It becomes clear that we need to introduce some kind of correction to the interplanetary speeds to be able to better predict the in situ speed from heliospheric images.

\begin{figure}
\epsscale{1.2}
\plotone{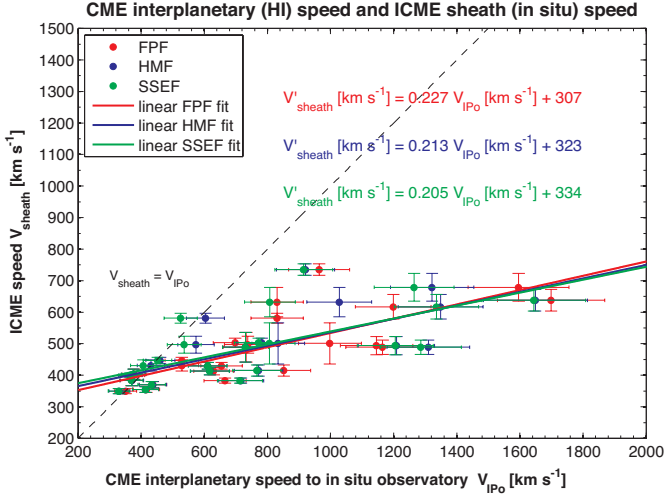}
\caption{Comparison between interplanetary CME speeds in the direction of the in situ observatory $V_{\mathrm{IPo}}$ (from geometrical modeling) and the observed proton bulk speeds $V_{\mathrm{sheath}}$ inside the ICME sheath regions. For each ICME, results for all three geometrical models are plotted (FPF red, HMF blue, SSEF green). Linear relationships derived for each of the three models are shown in the corresponding colors. The black dashed line indicates where the two speeds are equal.}\label{speed_scatter}
\end{figure}

Figure \ref{speed_scatter} presents the interplanetary speed ($V_{\mathrm{IPo}}$) in the direction of the in situ observatory against the ICME sheath region speed ($V_{\mathrm{sheath}}$). Separate linear fits for each geometrical model are shown in corresponding colors. These fits are very similar to each other. With the coefficients from these fits, we can apply an empirical correction to the speed $V_{\mathrm{IPo}}$ for each model, to get a better proxy for the speed of the in situ sheath region, which we call $V'_{\mathrm{sheath}}$ or a ``corrected'' speed:
\begin{gather}
V'_{\mathrm{sheath}}=0.227 V_{\mathrm{IPo}}+307, ~  ~ ~ ~ (\text{FPF model})\\
V'_{\mathrm{sheath}}=0.213 V_{\mathrm{IPo}}+323, ~  ~ ~ ~ (\text{HMF model})\\
V'_{\mathrm{sheath}}=0.205 V_{\mathrm{IPo}}+334, ~  ~ ~ ~ (\text{SSEF model})
\end{gather}
with both speeds given in \kmsec.  We use this formula to obtain corrected sheath speeds based on $V_{\mathrm{IPo}}$, and compare them again to the in situ observed $V_{\mathrm{sheath}}$. 

Returning to Figure~\ref{speed_bar}, the pink bars show the comparison $\Delta V=V'_{\mathrm{sheath}}-V_{\mathrm{sheath}}$, for the SSEF method only. It can be seen that the consistency with $V_{\mathrm{sheath}}$ greatly improves, for both slow and fast CMEs. On average, $<|\Delta V|>=50$~\kmsec~(FPF), 53~\kmsec~(SSEF) and 57~\kmsec~(HMF); see again Table~3 for a summary. For the whole dataset, for FPF,  the absolute difference $|\Delta V|< 100$~\kmsecc in 92 \% of cases (SSEF: 83 \%; HMF: 88\%).  This is an improvement over more than 200 \kmsecc compared to geometrical modeling without this empirical correction. Clearly, the caveat of this correction is that we derived it from the current set of events, and the results will change for a different set of CMEs. Nevertheless, it definitely results in an improvement in the prediction capability compared to using the geometrical models alone, and should also be used in future studies.

In summary, the average difference between CME speeds predicted from  geometrical modeling to in situ ICME sheath region speeds are, for all methods, using $V_{\mathrm{IP}}$, $<|\Delta V|>= 431$ \kmsec; using $V_{\mathrm{IPo}}$, $<|\Delta V|>=284$ \kmsec, and using $V'_{\mathrm{sheath}}$, $<|\Delta V|>= 53$ \kmsec. The differences between the different geometricals model are negligible, so none can be claimed as performing significantly better than the other.


\subsubsection{Interplanetary distance ranges}

Before we move on to discuss similar comparisons of ICME arrival times, we first need to clearly state the range of distances in the interplanetary medium that are covered by the \emph{STEREO/HI} observations.  Using J-maps created by the \emph{SATPLOT} software, we tracked every CME event out to about 30--40\degree elongation from the Sun. Rather than elongation, we need to know the corresponding distance range in AU to see how far from the Sun we have actually tracked the front of each CME. The average elongation value of the last HI data point of each CME is $34.9\pm 7.1$\degreee. The corresponding distance to the maximum elongation is  calculated by  $d_{\mathrm{p}}=(t_{\mathrm{p}}- t_{0}) \times V_{\mathrm{IP}}$. This means that we propagate the CME away from the Sun with the interplanetary speed of the model apex, from the launch time to the last time of HI observation ($t_\mathrm{p}$). For simplicity, we state the result as a mean and standard deviation for all CMEs for the SSEF model only, because it is intermediate between the FPF and HMF models. The average maximum distance of the CME apex from the Sun, over all CMEs, is  $<d_\mathrm{p}>=0.855 \pm 0.276$~AU (SSEF). The maximum value is 1.41 AU, and the minimum 0.48 AU. The maximum apex distance for some events is $> 1$~AU, especially for those cases where the CME flank hits the in situ spacecraft, which is situated near 1 AU. \\

\subsubsection{Prediction lead times}

We focus on connecting CMEs in different datasets, so we track the CMEs in the HI data for as long as possible. But we can also ask ourselves how useful the results will be for forecasting the in situ parameters. For this it is necessary to place the parameters that we have calculated from remote images, and their comparison to in situ observations, into the context of space weather prediction.

A very important parameter in this respect is the prediction lead time ($t_{\mathrm{lead}}$). This is the difference between the actual in situ arrival time of the ICME, and the time ($t_\mathrm{p}$) when a prediction for its arrival was issued. In our study, the latter is again the last point at which the event was observed by HI (i.e., the last point clicked in a J-map); of course here we are not discussing real time predictions but artificial ``predictions'' made long after the events have happened, derived from a pre-existing dataset. We define $t_{\mathrm{lead}}=t_{\mathrm{p}}-t_{\mathrm{insitu}}$, where $t_{\mathrm{insitu}}$ is the actual ICME arrival time.  With this definition, prediction times earlier than the actual arrival give a $t_{\mathrm{lead}}< 0$, as desirable when really ``predicting'' CMEs. Over our dataset, on average, $<t_{\mathrm{lead}}>= -26.4 \pm 15.3$~hours, with a minimum $t_{\mathrm{lead}}$ for any individual event of $-53.6$~hours ($> 2$ days), and a maximum of $+0,28$h. For the latter, $t_p$ was already later than the in situ arrival time. Hence, the results that we state for arrival times and speeds are valid for a prediction lead time of the order of one day, albeit with considerable scatter around this value. We also need to point out that we have used HI science data, which are of better quality than those available from \emph{STEREO} in real time.


\subsubsection{Arrival times}

\begin{figure*}
\epsscale{1.2}
\plotone{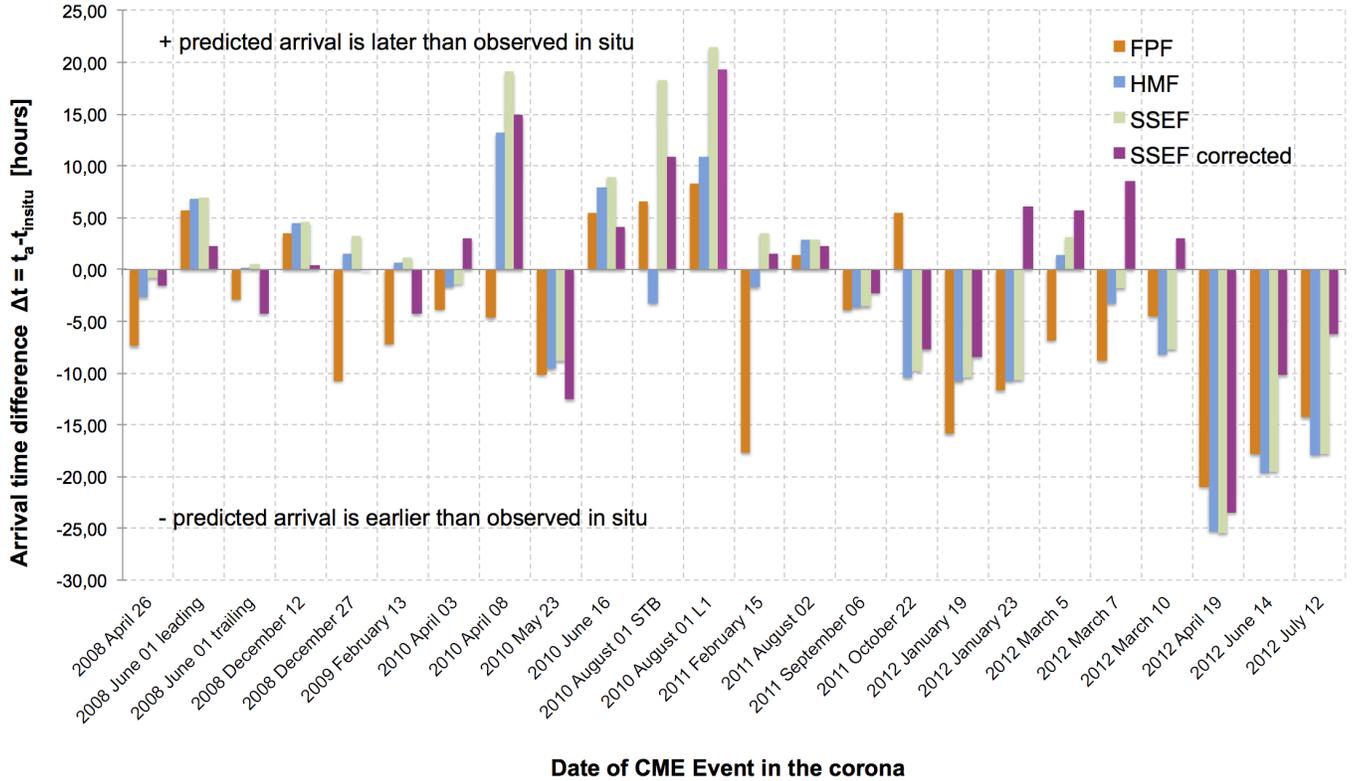}
\caption{Difference between CME arrival time as derived from geometrical modeling and actual ICME arrival time at the in situ observatory, for each CME event and each geometrical model (FPF: red; HMF: blue; SSEF: green). Pink bars indicate differences between corrected arrival times and the observed in situ arrivals ($\Delta t=t_{a;corr;v1}-t_{in situ}$). The corrected arrival times (for SSEF only) were calculated with the linear relationship shown in Figure~\ref{arrival_scatter}. }\label{arrival_bar}
\end{figure*}

Figure \ref{arrival_bar} is a bar chart showing, for all CMEs, the difference $\Delta t=t_\mathrm{a}-t_{\mathrm{insitu}}$ between predicted CME arrival time and the observed ICME arrival time taken from the in situ measurements. Here, $t_\mathrm{a}$ is the predicted arrival time from heliospheric imaging, derived by the following expression \citep[similar to][]{moe11,moe13},
\begin{equation}\label{eq:arrival}
t_\mathrm{a}=t_0+\frac{d_\mathrm{i}}{V_{\mathrm{IPo}}},
\end{equation}
where $t_0$ is the launch time, as discussed above, $d_\mathrm{i}$ is the heliocentric distance of the in situ spacecraft, and $V_{\mathrm{IPo}}$ the (assumed constant) CME speed in the direction of the in situ observatory. 

In Figure~\ref{arrival_bar}, values  of $\Delta t< 0$ refer to the situation where the predicted arrival time is before the in situ arrival time. This is the case in particular for the events after October 2011, on the right hand side of Figure \ref{arrival_bar}, which are all backsided events, from the HI observer point of view, as determined by the SSEF direction. Even so, the fastest CMEs after October 2011 have a reasonable $|\Delta t| < 10$~h, but this is mainly due to the fact that fast CMEs have a transit time of the order of 20--40 hours, compared to around 100 hours for slower events. In general, we find $\Delta t$ is of the order of 10\% of the total CME transit time. However, all three methods predict arrival times that are one day early for slow and backsided CMEs (e.g. 2012 April 19 with $V_{\mathrm{init}}=639$~\kmsec), consistent with theoretical expectations \citep{lug13}.

There is considerable scatter around $\Delta t \approx 0$, and for a specific event, different methods can give quite different arrival times. The maximum and minimum values of $\Delta t$ are about $\pm 1$~day, while the average absolute values of $<|\Delta t|>=8.5 \pm 5.4$h for the FPF model (SSEF: $8.4 \pm 7.2$h; HMF: $7.5 \pm 6.6$h; $8.1 \pm 6.3$h over all methods). Results are summarized in Table~3, including averages of signed values of $<\Delta t>$.

\begin{figure}
\epsscale{1.2}
\plotone{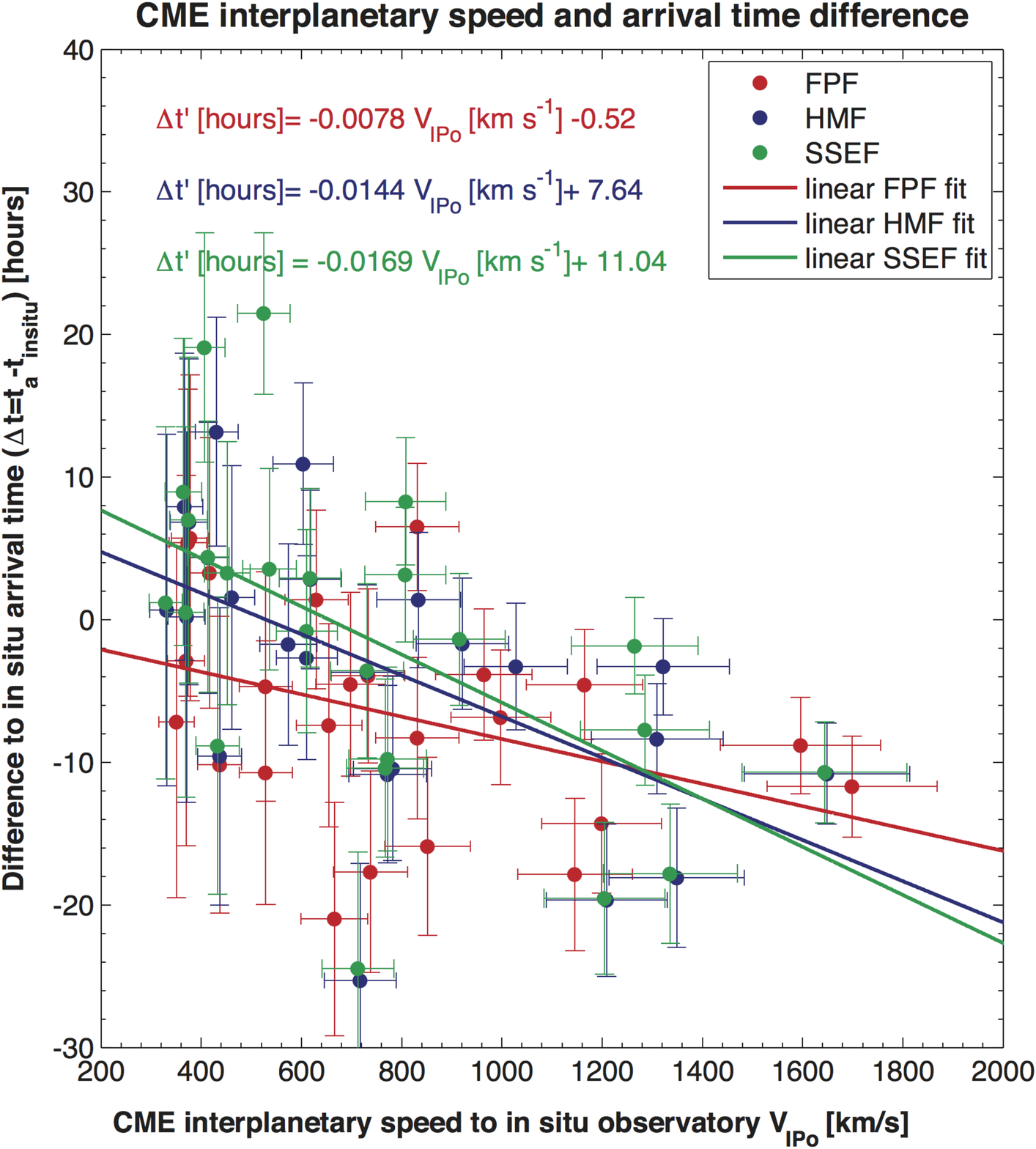}
\caption{Scatter plot of the difference between observed in situ ICME arrival times and predicted arrival times from geometrical modeling. Linear fits are shown as solid lines: FPF (red), HMF (blue), SSEF (green). Corresponding relationships are quoted on the figure in the respective color. }\label{arrival_scatter}
\end{figure}

We have derived corrections to the arrival times, using a similar approach as for correcting the speed predictions, but independent of the results for the speeds. Figure~\ref{arrival_scatter} is a scatter plot of the interplanetary speed $V_{\mathrm{IPo}}$ versus $\Delta t$, for each CME and for all three methods. Faster interplanetary speeds tend to be correlated with lower values of $\Delta t$, attributable again to the constant speed assumption used in geometrical modeling. Faster CMEs are predicted to arrive before they actually do, because modeling a fast and decelerating CME assuming a constant speed results in a much higher speed than the final in situ speed, as we have shown in the previous section. Similar to what was done in Figure~\ref{speed_scatter}, we fit the data points in Figure~\ref{arrival_scatter} for each model independently with a simple, linear relationship. The results of these fits are:
\begin{align}
\Delta t'=-0.0078 V_{\mathrm{IPo}} - 0.52 ~ ~ ~ (\text{FPF model}), \\
\Delta t'=-0.0144 V_{\mathrm{IPo}} +7.64 ~ ~ ~ (\text{HMF model}),\\
\Delta t'=-0.0169 V_{\mathrm{IPo}} +11.04 ~ ~ ~ (\text{SSEF model}),
\end{align}
with $\Delta t'$ in hours and $V_{\mathrm{IPo}}$ in \kmsec. This allows to make  an empirical correction (that we call ``version~1'') to the arrival time from our knowledge of $V_{\mathrm{IPo}}$ alone. To this end, we calculate a corrected arrival time by 
\begin{equation}
 t_{\mathrm{a;corr;v1}}=t_\mathrm{a}-\Delta t',
\end{equation}
using $\Delta t'$ from the fit for the corresponding geometrical model. The result of such a calculation for the SSEF model is shown using pink bars in Figure~\ref{arrival_bar}. On average, it improves the prediction performance by 2 hours. Comparing corrected arrival times to in situ arrival times results in $<|\Delta t|>= 5.6 \pm 4.1$h for the FPF model (SSEF: $6.8 \pm 6.0$h; HMF: $ 5.9 \pm 4.9$h). The average over all models is  $6.1 \pm 5.0$h, compared to 8.1 hours without the empirical correction (see Table~3). 

\subsubsection{Transit times} 

There is another way to calculate CME arrival times which performs slightly better than the straightforward arrival time calculation using geometrical modeling, as stated in Equation~(\ref{eq:arrival}). We have applied this new method, that we call ``version 2'', which also depends only on $V_{\mathrm{IPo}}$, to our list of events. We define a CME's transit time $TT$ as the time difference between its first appearance in coronagraph images ($t_{\mathrm{COR2}}$), and its actual in situ arrival time $t_{\mathrm{insitu}}$. This definition of $TT$ covers a distance range from $2.5~R_{\odot}$, which is the inner boundary of  \emph{STEREO/COR2}, to 1~AU or $\approx 215~R_{\odot}$.

\begin{figure}
\epsscale{1.2}
\plotone{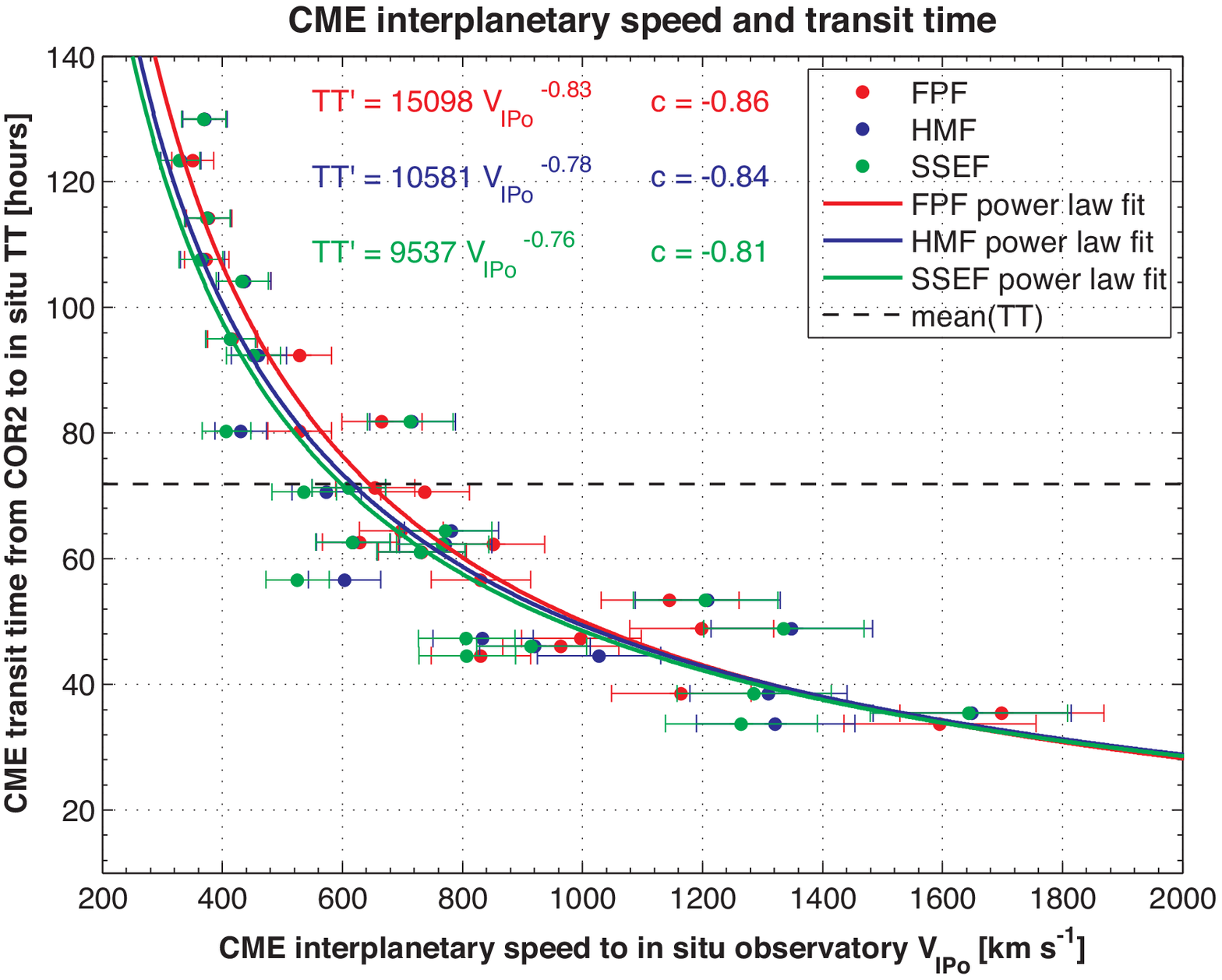}
\caption{CME interplanetary propagation speeds $V_{\mathrm{IPo}}$, in the direction of the in situ observatory, plotted against CME transit times $TT$. For each event, $V_{\mathrm{IPo}}$ from each of the three geometrical models are given: FPF (red), HMF (blue), and SSEF (green). Power law fits for each model are plotted as solid lines, with the function and the correlation coefficent ($c$) for each method quoted on the image. The black dashed line shows the mean transit time for the CMEs in our sample, $<TT> = 71.9$ hours.  }\label{arrival_transit}
\end{figure}

Figure~\ref{arrival_transit} shows a relationship between $TT$ and interplanetary speeds ($V_{\mathrm{IPo}}$) for all CMEs in our dataset. For each geometrical model, we fit a power law to the data, quoting the results on the plot. For SSEF, the resulting power law is: 
\begin{equation}
TT'=9537 \times V_{\mathrm{IPo}}^{ -0.76} ~  ~ ~ ~ (\text{SSEF model},)
\end{equation}
with $V_{\mathrm{IPo}}$ given in \kmsec, and $TT'$ in hours. Similar relationships have been shown for example by \cite{sch05} and \cite{vrs07}, but using CME coronagraph speeds, whereas we show interplanetary speeds. They include to some extent the CME propagation through the background solar wind. The average transit time in our dataset is around 72 hours (or 3 days), which compares well to the classic \cite{bru98} 80 hour rule for average CME transit times. These power laws allow us to predict the arrival time, based solely on the interplanetary speed in direction of the in situ observatory $V_{\mathrm{IPo}}$, by

\begin{equation}
t_{\mathrm{a;corr;v2}}=t_{0} + TT'.
\end{equation}

For simplicity, we do not show resulting arrival time differences $\Delta t$ for each event, but only summarize the resulting values. For this method, the difference between predicted and in situ arrival times is $<|\Delta t|>= 5.7 \pm 3.8$h for the FPF model (SSEF: $7.7 \pm 6.1$h; HMF: $6.6 \pm 5.3$h). The average for all models is $6.6 \pm 5.1$ hours, almost as good as for the empirically corrected arrival times (6.1 hours).

\subsubsection{CME speed and ICME magnetic field}

The last step in our analysis concerns the relationship between the maximum magnetic field $B_{\mathrm{max}}$ in the full ICME interval (column~8 in Table 2), consisting of shock, sheath and ejecta parts, and the interplanetary speed $V_{\mathrm{IPo}}$ in the direction of the in situ observatory, as derived from HI geometrical modeling. CMEs that have higher speeds close to the Sun are known to be correlated with stronger magnetic fields at in situ spacecraft \citep[e.g.][]{lin99,yur05}. This is thought to arise from the role played by magnetic reconnection during CME eruption, namely through a possible formation of their interior magnetic flux rope \citep[e.g.][]{qiu05,qiu07,moe09}. Another explanation is that faster CMEs lead to stronger compression of the ambient solar wind, which results in a higher magnetic field strength in the ICME sheath region \citep{liu08}. Around half of the ICMEs in our dataset reach their $B_{\mathrm{max}}$ in the sheath and the other half in their interior magnetic ejecta, so both effects can be expected to play a role in our analysis. In the same way as we have done for the transit times above, we can extend such previously found relationships by comparing interplanetary rather than coronal CME speeds to the in situ magnetic field magnitude. 

\begin{figure}
\epsscale{1.2}
\plotone{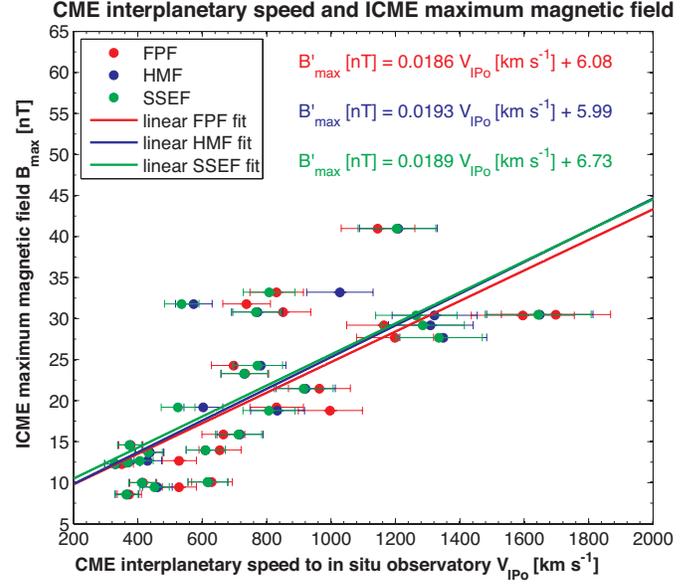}
\caption{Relationship between the in situ observed maximum magnetic field $B_{max}$ in the full ICME interval (shock, sheath, and ejecta) and the interplanetary CME speed in direction of the in situ observer ($V_{\mathrm{IPo}}$). Formulas for linear fits to the data, for each model, are given in corresponding colors. }\label{magnetic_relation}
\end{figure}

Figure~\ref{magnetic_relation} shows, as a scatter plot, that stronger ICME maximum magnetic fields are correlated with faster interplanetary CME speeds. Again, we perform independent, linear fits for each model, which, for SSEF, results in
\begin{equation}
B'_{\mathrm{max}}=0.0189 V_{\mathrm{IPo}}+6.73,
\end{equation}
with $B_{\mathrm{max}}'$ in nT, and $V_{\mathrm{IPo}}$ in \kmsec. The average, absolute difference $\Delta B=B'_{\mathrm{max}}-B_{\mathrm{max}}$ between predicted and observed maximum magnetic fields, based on the speeds from FPF, is $<|\Delta B|>=4.8$~nT (SSEF: 4.7 nT; HMF: 4.4 nT). Use of the resulting linear fits for $B_{max}'$ for each model results in a successful prediction of $B_{\mathrm{max}}$ to within $\pm 5$~nT  for 58~\% of events with the FPF method (SSEF: 67\%; HMF 71\%). The same numbers for predictions within $\pm 10$~nT are 88 \% for FPF  (SSEF: 88\%; HMF 92\%). In summary, for around 2 out of 3 events in our sample, the maximum magnetic field in the ICME interval could be predicted within a precision of 5 nT, solely based on the interplanetary CME speed.

\section{Conclusions}
We have presented a study in which we have connected white light to in situ observations of a set of 22 near-equatorial CMEs. Their parameters  cover a wide range of initial and propagation speeds (260 to 2715 \kmsec), and a variety of principal CME directions with respect to the HI observer (40\degree to 170\degree in heliocentric longitude). This is the first study to do such an analysis of such a large and diverse set of events. Here, we summarize our results and their implication for space weather prediction and the evolution of CMEs between the Sun and 1~AU separately, emphasizing our main conclusions in italics.

First, concerning space weather prediction, we have mainly tested methods for predicting CME directions, speeds and arrival times based on J-maps produced using observations from a single-spacecraft heliospheric imager instrument. These data were provided by HI on \emph{STEREO-A/B}, which, for the CMEs under study, were positioned well outside the Sun-Earth line, on heliocentric longitudes between 30\degree and 120\degree away from the Earth. 

\begin{enumerate}
\item \textbf{Event selection}: We have selected CME events with clear interplanetary CME signatures (shock, sheath and ejecta) in \WIN and \STB in situ observations near 1 AU, and which we could easily track in the HI J-maps. This set of CMEs covers a wide range of initial speeds, so it is suitable for statistical analysis, although there are actually many more CMEs in the data during this time range. The events in our list occurred during the rise of solar cycle 24, between 2008-2012, and \emph{STEREO} progressed towards and beyond opposition during this time. Therefore, our dataset contains mostly slow frontsided and fast backsided CMEs. Our results must be considered carefully with this in mind, and are likely to be different for different datasets.

\item  \textbf{Tracking CMEs and prediction lead times:}
We tracked CMEs as far as possible in the J--maps, resulting in an average track length of $34.9\pm7.1$\degree  elongation. Depending on the CME speed, its direction and the model applied, this corresponds to different distances each CME has traveled away from the Sun at the time of the last HI observation. This distance is on average $0.86 \pm 0.28$ AU, thus we have tracked most CMEs almost up to 1 AU. The prediction lead times also depend on several of the aforementioned parameters, and are on average $-26 \pm 15$~hours, about $- 1$~day. This is considerably longer than those provided by a solar wind monitor at L1  (on the order of 1 hour, e.g.\ the current \emph{ACE} mission), but definitely shorter than lead times using corongraphs only \citep[on the order of CME transit times of a few days, e.g.][]{kil12b}. The work presented here should rather be seen as an attempt to establish connections between CME-related phenomena detected in different data sets (coronagraph, heliosphere, in situ), which is strongly desirable for benchmarking various empirical or numerical CME prediction models.

\item \textbf{CME directions}: The three different geometrical models used in this study do not provide consistent CME directions, with differences of up to 50\degree in heliocentric longitude for fast CMEs. For this reason, we derived the direction from multi-viewpoint applications of the croissant model to coronagraph images as a well constrained reference for the initial CME direction. The FPF direction was shown to be, on average, consistent to within a few degrees with the croissant direction. This means that \emph{if CMEs propagate radially away from the Sun above 15 solar radii, the FPF method is superior over the others in deriving the CME principal direction}. However, there is also considerable scatter, of the order of $\pm 30$\degree heliocentric longitude around the average direction. The directions provided by the SSEF and HMF models, which assume an extended CME front, differ by about $30$\degree from the croissant direction. These methods show a bias in the direction to values further away from the HI observer, with larger differences for faster CMEs, confirming theoretical predictions by \cite{lug13}.

\item \textbf{CME arrival time and speed predictions:} Using geometrical modeling 
to predict CME speed and arrival time, with a lead time of roughly 1 day, results in average absolute differences from observed ICME sheath speeds and arrival times of $|\Delta V|=284 \pm 288$~\kmsecc and $|\Delta t| =  8.1 \pm 6.3$h, respectively, averaged over all three methods. The arrival time difference given by the root-mean-square method is  $\Delta t_{RMS} =10.3$h. Conserving the sign of the differences, similar numbers for calculated minus observed speeds and arrival times are $\Delta V=275 \pm 297$~\kmsecc and $\Delta t =  -3.8 \pm 9.6$h, respectively. \emph{In summary, predicted ICME speeds at 1 AU are, on average, overestimated, and consequently, predicted arrival times are too early compared with the in situ observations}. Correlating the interplanetary CME speed in the direction of the in situ observatory with the in situ speeds, and, independently, with the differences in observed and predicted arrival times, we obtained new empirical relationships between these variables, based on linear fits. These allow us to predict the CME speed and arrival time with better accuracy for, at least, the present dataset. \emph{Applying these empirical corrections in the prediction of CME speeds and arrival times, 
$|\Delta V|=53 \pm 50$~\kmsec ($\Delta V=0 \pm 73$~\kmsec), and $|\Delta t| =  6.1 \pm 5.0$h ($\Delta t =  0.0 \pm 7.9$h, $\Delta t_{RMS} = 7.9$h), averaged over all events and methods.} More specifically, 88\% (or roughly 9 out of 10) of all in situ ICME sheath region speeds in our dataset were predicted to within $|\Delta V| < 100$~\kmsec. In  71\%  of cases (or roughly 3 out of 4), ICME arrival times are predicted within $|\Delta t| < 8$~hours. Moreover, we quantified the relationship between the maximum magnetic field $B_{max}$ in the ICME and the interplanetary propagation speed, which lets us predict $B_{max}$ to within $\pm 5$~nT in 2 out of 3 cases.

\item \textbf{Comparison to stereoscopic modeling}: \cite{col13} found similar results for CME arrival time and speed prediction using the results of methods applied to two heliospheric imager instruments, such as fitting the stereoscopic croissant model to \emph{STEREO/COR2/HI1/HI2} and \emph{SOHO/LASCO} images, for heliocentric distances up to about 0.9 AU. Similar to our study, they also applied corrections for the apex and flank parts of the CMEs impacting the in situ observatory. They found that for 78\% of cases, $\Delta t < 6$~hours, and for 55\% of CMEs, $\Delta V < 100$~\kmsec, although for all events $\Delta V < 140$~\kmsec.  \cite{col13} fitted the height-time data in several different ways, and they found a linear fit between $50$~$R_{\odot}$ (0.23 AU) and 1~AU and thus assuming a constant speed to produce the best predictions, even though they also concluded that CMEs clearly decelerate. It is interesting to note that an equivalent assumption of constant speed between 0.07 to 0.85 AU in our geometrical modeling methods yields a similar level of predictive performance.
\end{enumerate}


Second, we summarize our results in terms of the physical evolution of CMEs between the Sun and 1 AU. 
In general,  we find that, using single-spacecraft HI observations, it is much more difficult to generalise the behavior of CMEs in terms of such properties as propagation direction,  speed variation and global shape evolution when compared  to the use of stereoscopic HI measurements \citep[e.g.][]{liu10, lug10, liu11, col13, mis13, liu13, dav13}. 

\begin{enumerate}
\item \textbf{CME radial propagation:} A major unsolved question is whether CMEs propagate radially away from the Sun, or undergo any significant change of direction as they travel through the interplanetary (IP) medium. CMEs are known to be deflected in the corona by coronal holes \citep{gop09} and possibly through CME-CME interaction in IP space \citep{lug12}. \cite{liu13} and \cite{dav13} showed the CME direction for a few events to stay within about $\pm 15$\degree (heliocentric longitude) from the corona to the IP medium. This can still be considered as consistent with a radial propagation. With our own dataset, CME deflections are difficult to assess without further analysis. As discussed above, the HI methods can give very different answers for a single event - \emph{CME principal directions are not well defined using single-spacecraft HI methods.} Note also that we assume constant direction in IP space in contrast to the stereoscopic methods, so we can only assess deflections by comparing directions derived from coronagraph and HI measurements. However, there is considerable scatter between these initial and IP directions (of the order of $\pm 30$\degreee, about twice the level derived from stereoscopic methods), pointing indeed to the possibility that some CMEs may significantly change in direction between the corona and the IP medium. While the croissant model constrains the CME direction well close to the Sun, there is no straightforward way to constrain the CME direction using single-spacecraft in situ data, and thus we leave further analysis for future work.

Our results on CME direction in IP space clearly point out that its calculation is always influenced by the strongly idealized shapes of the CME front which we assume in the first place. The geometrical definitions we use make it possible to describe CME front shapes analytically, and form very useful tools. However, our study casts some doubts on their use in defining a CME's central propagation direction in the interplanetary medium, because its calculation depends so strongly on the assumption of its frontal geometry. Consequently, we propose that \emph{future work should quote ranges of heliocentric longitude that will be affected by a CME, rather than a central direction}. This would be especially helpful when describing distorted or asymmetric CME front shapes \citep[e.g.][]{sav10,moe12}.

\item \textbf{CME speed profiles}: CMEs are well known to decelerate in the solar wind out to 1 AU, mainly due to a force equivalent to aerodynamic drag \citep[e.g.][]{gop01, kil12b, vrs13,liu13}, and we can see this clearly in our data.  We can compare stepwise a CME initial speed ($< 15~R_\odot$, or $< 0.07$ AU) to its average IP propagation speed (from $\approx 0.07$ to $0.86 \pm 0.28$ AU), and to the speed of the ICME sheath region observed in situ near 1~AU. FPF and SSEF suggest that CMEs decelerate from the corona to IP space, while HMF yields a constant speed. CMEs are clearly slower when observed in situ near 1 AU than in IP space, and the in situ sheath region speeds can be reasonably well predicted by multiplying the IP speeds with $\approx 0.2$ and adding $320$~\kmsec, which is independent of the model used. The speed of $320$~\kmsecc is reminiscent of the slow solar wind, and  this relationship is also consistent with early work by \cite{lin99}. However, we cannot definitely say where most of the deceleration occurs, because of our assumption of constant IP speed. Recent analyses with stereoscopic methods by \cite{liu13}  showed that much of the deceleration of fast CMEs occuring up to about 80 solar radii (0.37 AU), but there is probably no general distance by which the deceleration ceases, as a case has been found where a CME almost does not decelerate out to 1 AU \citep{moe10}.
  
\item \textbf{Evolution of the global CME shape}: We expected that there would be clear differences concerning the prediction performance of different geometrical models, giving us hints which one better describes the CME front shape in the plane perpendicular to the HI images. In particular, we expected the SSEF model, as it is the most mature, with a well defined CME width, to perform best. But surprisingly,  we did not find any significant difference in its performance for predicting CMEs compared to the other models. We think that this is caused by their assumptions of constant speed and constant direction, and their high sensitivity to a violation of these assumptions \citep{lug13}. In summary, \emph{the current state of the art of geometrical modeling of CMEs with single-spacecraft instruments (i.e.\ fitting methods) in comparison to single-spacecraft in situ data precludes inferences to be made regarding the large--scale geometry of CME fronts in planes perpendicular to the HI images, such as the ecliptic plane.}
\end{enumerate}

We conclude that predicting CME speeds and arrival times with heliospheric images gives more accurate results than using projected initial speeds from coronagraph measurements. These improvements are on the order of 12 hours for the arrival times \citep{col13}, and our results are consistent with those found with other space weather models in the \ST era \citep[see also][]{gop13,mis13,mis14}. Independent of the specific methods used,  we can derive an average of the CME interplanetary propagation speed when we track a CME out to 1 AU. This average speed includes to some extent the background solar wind, which is known to play a significant role in modulating CME propagation \citep[e.g.][]{gop01,vrs07, tem11, kil12b}. The same is true for interacting CMEs, although it can be very difficult to differentiate and decipher different density tracks in HI when CMEs are launched close together in time and space \citep{har12, web13}. 

It is clear that the information contained in heliospheric images does indeed improve space weather prediction. However, the compromise between prediction accuracy ($\Delta t$) and prediction lead times ($t_{lead}$) needs to be better studied. Future modeling should thus focus on modeling CME tracks in J-maps for elongations from the Sun of $< 35$\degreee, which will result in longer prediction lead times than one day in the current study. It needs to be assessed how the predictions for speed and arrival time become less accurate as the prediction lead time is increased, and the best balance for space weather prediction purposes needs to be found. 

Another possibility for future work is to actually include deceleration into the geometrical fitting methods for single-spacecraft HI instruments. This is expected to further improve the accuracy of predicting CME arrival times and speeds through use of a more realistic approach than the current assumption of constant speed. Such  an approach, applied to the different model geometries, could provide clues to the CME global front shape, in particular when constrained by multi-point in situ measurements.

Another solution would be to send a mission equipped with a heliospheric imager in a polar orbit around the Sun to look down onto the ecliptic plane \citep{lie08}. In this way, information on the global shape of a CME in the ecliptic, as it approaches Earth, would be revealed, and the CME propagation characteristics could be provided by the geometrical models as used in our paper. However, for effectively predicting geo-effectiveness it is also necessary to know the components of the interplanetary magnetic field, which cannot be derived from white-light images, and require new ideas like the Faraday-rotation technique \citep[e.g.][]{xio13}.
Moreover, the \emph{Solar Orbiter} \citep[e.g.][]{mue13} and \emph{Solar Probe Plus} missions are currently under development and will approach the Sun closer than ever before by the end of the decade. The  results and methods  presented in this paper  may provide both the scientific background and tools for analyzing the data  from these exciting future  missions.


\begin{table*}[t]\label{tab:imaging}
\begin{center}
\caption{Imaging: Croissant and geometrical modeling (SSEF) results}
\begin{tabular}{ccrrrrrrcc}
\tableline\tableline
(1) & (2) & (3) & (4) & (5) & (6) & (7) &(8) & (9) &(10)\\
CME & \emph{t$_{COR2}$} & $\Phi_{\mathrm{init;Earth}}$ & \emph{V$_{\mathrm{init}}$} & $\Phi_{\mathrm{IP;Earth}}$ & $\Phi_{\mathrm{IP;HI}}$& \emph{V$_{\mathrm{IP}}$} & \emph{V$_{\mathrm{IPo}}$} & \emph{t$_a$} & spacecraft\\
\tableline
1 &	2008 Apr 26  14:53 &  -20&	523&-45&-70& 656&611& 2008 Apr 29 13:21 &STEREO-B\\
2 &	2008 Jun 01 21:23&-37&260 &	-15&-44 &384& 375	& 2008 Jun 06 22:34&	STEREO-B\\
3 &	2008 Jun 02 02:07&	\nodata\tablenotemark{a}&	\nodata &-44&-73&398& 369&2008 Jun 07 12:39	&STEREO-B\\
4 &	2008 Dec 12 07:37&8&	497&10&55&421&414&2008 Dec 16 11:01&Wind\\
5 &	2008 Dec 27 05:23&  -32& 405   &-79     &-122&603&452&2008 Dec 31 05:01	&STEREO-B\\
6&	2009 Feb 13 06:37&	\nodata\tablenotemark{b}&	\nodata      &	-75 &	-119 &	396	&329&	2009 Feb 18 11:11&Wind\\
7 &	2010 Apr 03 09:54&	4  &	829  &	-19	&-86	&991	&915&	2010 Apr 05 06:35	&Wind\\
8&	2010 Apr 08 03:54&	3  &	511	 &  -34&	-102&	555&	407&	2010 Apr 12 07:19	&Wind\\
9&	2010 May 23 17:39&	10 & 381	  &    8	&-63&	440&	433	&2010 May 27 17:02&	Wind\\
10&	2010 Jun 16 11:24&   -16  & 297 &	13&	-62&	376&	364&	2010 Jun 21 08:00&	Wind\\
11&	2010 Aug 01 08:24&  -28 & 1160 &	-43	& -122 &	980	&808	&   2010 Aug 03 13:18&	STEREO-B\\
12&	2010 Aug 01 08:24&  -28 & 1160 &	-43	&-122 &	980	&525	&     2010 Aug 04 14:34&	Wind\\
13&	2011 Feb 15 02:09&	23 & 557	   & -40	&-127&	867	&536	&     2011 Feb 18 04:21&	Wind\\
14&	2011 Aug 02 06:39&  -29 & 1050 &	-25	&-125&	714	&617	&     2011 Aug 05 00:14&	Wind\\
15&	2011 Sep 06 22:39&    34 &	 1160 &	-24	&-127&	838	&731	&     2011 Sep 09 08:11&	Wind\\
16&	2011 Oct 22 01:09&    19 & 692	   & -15	&-120&	813	&772	&     2011 Oct 24 07:52&	Wind\\
17&	2012 Jan 19 15:09&   -37 &	 1335  &	-36	&-144&	1097&	767	  &   2012 Jan 21 19:03&	Wind\\
18&	2012 Jan 23 03:09&     21 & 1708  &	-33	&-141&	2181&	1644 &	2012 Jan 24 03:54&	Wind\\
19&	2012 Mar 05 04:09&  -53 & 974	    & -41	&-150&	1347&	807	  &   2012 Mar 07 06:37&	Wind\\
20&	2012 Mar 07 01:39\tablenotemark{c}&  -35 & 2585  &	-42	&-151&	2202&	1265 &	2012 Mar 08 08:34&	Wind\\
21&	2012 Mar 10 17:54&    27 & 1265  &	-6	&-116&	1297&	1286 &	2012 Mar 12 00:44&	Wind\\
22&	2012 Apr 19 16:24&  -26 & 639	    & -20	&-133&	785	 &     713	 &    2012 Apr 22 01:46&	Wind\\
23&	2012 Jun 14 14:09&	 0  & 1102  &	-28	&-145&	1453 &	1205 &	2012 Jun 16 00:02&	Wind\\
24&	2012 Jul 12 16:45&	-1 &	1277  &  -22	& -142&	1486 &	1336 &	2012 Jul 13 23:50& 	Wind\\
\tableline
\end{tabular}\tablenotetext{1}{This event is the core or trailing part of the CME on June 1, so it is not fitted with the croissant model.}
\tablenotetext{2}{This CME was too faint to be fitted with the croissant model.}
\tablenotetext{3}{This is the 2nd CME of a double eruption of the same active region, with another CME first observed at 00:39 UT in COR2A. We quote the croissant results for the 2nd event because it is mainly directed in the ecliptic plane and thus more likely to impact Earth, whereas the first CME is directed mainly northward.}
\tablecomments{Explanation of parameters: (1) Number of CME event. (2) Date and time (UT) of the first image in COR2A when the CME is observed. (3) The initial direction (in degrees) in heliocentric longitude (\emph{Heliocentric Earth Equatorial} or \emph{HEEQ} coordinates) of the CME from croissant modeling (2.5-15.6~$R_{\odot}$). Earth is at 0\degree longitude, angles $> 0$\degree corresponds to solar west. (4) The initial speed of the CME from croissant modeling, in \kmsec. (5) The interplanetary direction (in degrees) of the CME in heliocentric longitude (close to but not exactly in the solar equatorial plane, see Section 2.2), from geometrical SSEF modeling with 45\degree half width (Earth at 0\degreee). (6) The interplanetary direction of the CME apex in degrees, measured from the HI observer (at 0\degreee). (7) The CME interplanetary propagation speed, for the apex of the front (for the SSEF model), in \kmsec.  (8) The speed of the point along the CME front that travels towards the in situ spacecraft (for the SSEF model), in \kmsec. (9) Date and time (UT) of the predicted arrival at the in situ spacecraft, from SSEF modeling. (10) Name of the in situ spacecraft for which the predicted arrival time $t_a$ is calculated. }
\end{center}
\end{table*}


\begin{table*}[t]
\begin{center}
\caption{In situ ICME parameters}
\begin{tabular}{ccrrrrrrrr}
\tableline\tableline
(1) & (2) & (3) & (4) & (5) & (6) & (7) &(8) & (9) &(10)\\
CME & spacecraft & $d_i$ & $\Phi_{\mathrm{IP;insitu}}$ &\emph{t$_{\mathrm{insitu}}$} & \emph{V$_{\mathrm{sheath}}$} & \emph{N$_{\mathrm{sheath}}$} & \emph{B$_{\mathrm{max}}$} & min \emph{B$_z$} & min Dst\\
\tableline
1 & STEREO-B	& 1.0280 & 21 &2008 Apr 29 14:10  & 430$\pm$11 & 16$\pm$6  &	14.0 &	-9.5	 & \nodata\\
2 & STEREO-B	& 1.0542 & -11 &2008 Jun 06 15:35  & 403$\pm$16  & 15$\pm$7  &	14.6 &	-8.6 & \nodata\\
3 & STEREO-B	& 1.0548 & 19 &2008 Jun 07 12:07 &	384$\pm$17  & 13$\pm$7	&12.5	&-11.3	&\nodata\\
4 & Wind& 0.9840 & -10 &2008 Dec 16 06:36 &355$\pm$9	&16$\pm$4	&10.0	&-7.6	&\nodata\\
5 &  STEREO-B & 1.0263 & 34 &2008 Dec 31 01:45 &	447$\pm$10	&7$\pm$3	&9.5	&-6.7	&\nodata\\
6 & Wind	& 1.0023& 28 & 2009 Feb 18 10:00 &350$\pm$8	&22$\pm$4	&12.3	&-9.4	&\nodata\\
7 & Wind	& 1.0004& 19&2010 Apr 05 07:58 &735$\pm$18	&10$\pm$2	&21.5	&-14.6	&-81\\
8 & Wind	& 1.0021& 34&2010 Apr 11 12:14 &431$\pm$18	&10$\pm$1	&12.7	&-8.6	&-51\\
9&Wind	& 1.0132& -8&2010 May 28 01:52&	370$\pm$10	&19$\pm$4	&13.7	&-12.9	&-85\\
10&Wind	& 1.0161& -13&2010 Jun 20 23:02 &	400$\pm$6	&8$\pm$3	&8.6	&-2.8	&\nodata\\
11&STEREO-B	& 1.0604& -28 &2010 Aug 03 05:00&	632$\pm$47	&4$\pm$4	&33.2	&-30,2	&\nodata\\
12&Wind	& 1.0146& 43 &2010 Aug 03 17:05&	581$\pm$16	&10$\pm$2	&19.2	&-11.2	&-67\\
13&Wind	& 0.9881& 40 &2011 Feb 18 00:48&	497$\pm$27	&25$\pm$11	&31.8	&-24.3	&\nodata\\
14&Wind	& 1.0145& 25 &2011 Aug 04 21:18&	413$\pm$12	&6$\pm$1	&10.1	&-8.1	&-107\\
15&Wind	& 1.0072& 24 &2011 Sep 09 11:46&	489$\pm$47	&12$\pm$14	&23.3	&-21.4	&-69\\
16&Wind	& 0.9946& 15 &2011 Oct 24  17:38&	503$\pm$15	&26$\pm$4	&24.3	&-22.1	&-132\\
17&Wind	&0.9841 & 36 &2012 Jan 22  05:28&	415$\pm$18	&26$\pm$17	&30.8	&-27.9	&-69\\
18&Wind	& 0.9844& 33&2012 Jan 24  14:36&	638$\pm$34	&8$\pm$2	&30.5	&-15.7	&-73\\
19&Wind	& 0.9924& 41 &2012 Mar 07 03:28&	501$\pm$65	&14$\pm$5	&18.8	&-18.2	&-74\\
20&Wind	& 0.9927& 42&2012 Mar 08 10:24&	679$\pm$44	&12$\pm$4	&30.4	&-18.4	&-131\\
21&Wind	& 0.9938& 6 &2012 Mar 12 08:28&	489$\pm$23	&24$\pm$9	&29.2	&-23.6	&-50\\
22&Wind	& 1.0055& 20 &2012 Apr 23 02:14&	383$\pm$8	&24$\pm$7	&15.9	&-15.3	&-108\\
23&Wind	& 1.0160& 28 &2012 Jun 16 19:34&	494$\pm$29	&50$\pm$24	&41.0	&-21.0	&-71\\
24&Wind	& 1.0165& 22 &2012 Jul 14 17:38&	617$\pm$39	&16$\pm$6	&27.7	&-18.3	&-127\\
\tableline
\end{tabular}\tablecomments{Explanation of parameters: (1) Number of CME event. (2) The spacecraft which detected the ICME, using data from the \emph{PLASTIC} (plasma) and \emph{IMPACT} (magnetic field) instruments on \emph{STEREO-B}, and \emph{SWE} (plasma) and \emph{MFI} (magnetic field) on Wind. (3) Heliocentric distance of the in situ spacecraft at ICME arrival, in AU. (4) Difference between the interplanetary CME direction (heliocentric longitude, from the SSEF model) and the HEEQ longitude of the in situ spacecraft, in degree (Earth at 0\degreee). Small angles indicate central hits, larger angles flank hits. For angles $>0$\degree ($< 0$\degreee), the in situ observer is west (east) of the CME apex derived from SSEF. (5) The date and time (UT) of the in situ detection of the shock, or a significant increase in density ahead of a magnetic structure, if no shock is present. (6) The mean proton bulk speed in the ICME sheath region, and its standard deviation, in \kmsec. (7) The mean proton density in the sheath region, and its standard deviation, in cm$^{-3}$. (8) The maximum magnetic field (nT) in the ICME, including the sheath and ejecta intervals. (9) The minimum value of $B_Z$ (nT) in the ICME, the component of the magnetic field normal to the ecliptic plane (for Wind) or the solar equatorial plane (for \emph{STEREO-B}).  (10) The minimum value of the $Dst$ index (nT), provided by Kyoto, during the geomagnetic storm following the arrival of the ICME at Earth. Values $> -50$~nT and events directed towards \emph{STEREO-B} are ignored. }
\end{center}
\label{tab:insitu}
\end{table*}

\begin{table*}[t]
\begin{center}
\caption{Performance of HI geometrical models in connecting remote observations of CMEs to in situ data at 1 AU} 
\begin{tabular}{lccrrrr}
\tableline\tableline
method & variable & unit & FPF & SSEF & HMF & average \\
\tableline
Speed of model apex &$\Delta V$                                  & \kmsec &   298 $\pm$ 296   	& 443 $\pm$ 445      & 541 $\pm$ 567     & 427 $\pm$   455\\
Speed in direction of in situ observatory &$\Delta V$    &\kmsec &  298 $\pm$ 296 	& 252 $\pm$ 302        & 274 $\pm$ 304    & 275   $\pm$ 297 \\
Speed with empirical correction &$\Delta V$                  &\kmsec & 	0 $\pm$ 69           & 0 $\pm$ 79             &  0 $\pm$ 74          &  0  $\pm$ 73\\
Speed of model apex &$|\Delta V|$                                & \kmsec &  303 $\pm$ 290     & 446  $\pm$ 442 & 544 $\pm$ 564    & 431 $\pm$ 451\\
Speed in direction of in situ observatory &$|\Delta V|$  &\kmsec & 303 $\pm$ 290 	& 267  $\pm$  289& 282 $\pm$ 296    & 284 $\pm$ 288 \\
Speed with empirical correction &$|\Delta V|$               &\kmsec & 	50 $\pm$ 47        & 57  $\pm$  53   & 53 $\pm$ 51        & 53 $\pm$ 50\\
\tableline
Arrival time in direction of in situ observatory &$\Delta t$                 &hours &  -6.7  $\pm$    7.6  & -1.4    $\pm$   11.1  & -3.3  $\pm$  9.5 & -3.8 $\pm$  9.6 \\
Arrival time with empirical correction (=version 1) &$\Delta t$          & hours  & 0.0 $\pm$ 7.0     & 0.0 $\pm$ 9.2                  & 0.0 $\pm$ 7.8        &  0.0 $\pm$ 7.9 \\
Arrival time with transit time relationship (=version 2) &$\Delta t$    & hours & -1.9 $\pm$ 6.6	  &  1.2 $\pm$ 9.8                & -2.0 $\pm$ 8.3     & -0.9 $\pm$ 8.4\\
Arrival time in direction of in situ observatory & $|\Delta t|$               &hours &  8.5 $\pm$ 5.4      & 8.4 $\pm$ 7.2  & 7.5 $\pm$ 6.6  & 8.1 $\pm$ 6.3 \\
Arrival time with empirical correction (=version 1) &$|\Delta t|$        &hours & 5.6 $\pm$ 4.1       & 6.8 $\pm$ 6.0 & 5.9 $\pm$  4.9 &  6.1 $\pm$ 5.0\\
Arrival time with transit time relationship (=version 2) &$|\Delta t|$ & hours & 5.7 $\pm$ 3.8      & 7.7 $\pm$ 6.1 & 6.6 $\pm$ 5.3 & 6.6 $\pm$ 5.1\\
\tableline
\end{tabular}
\tablecomments{Explanation: For each method, the differences $C-O$ (calculated $-$ observed) between the parameters from geometrical modeling and the corresponding in situ parameters is given. The speed and arrival time from geometrical modeling are compared to the average proton bulk speed of the ICME sheath region, and the in situ arrival time of a shock or significant density jump, respectively. The numbers in the table correspond to the variables $\Delta V$ and $\Delta t$ in the text, and errors quoted correspond to $\pm 1$ standard deviation. }
\end{center}
\label{tab:performance}
\end{table*}

\acknowledgments

This research was supported by a Marie Curie International Outgoing Fellowship within the 7th European Community Framework Programme.  C.M. and T.R. thank the Austrian Science Fund (FWF): [P26174-N27]. M.T. was also supported by the Austrian Science Fund (FWF): V195-N16. The presented work has received funding from the European Union Seventh Framework Programme (FP7/2007-2013) under grant agreements no. 263252 [COMESEP] and no. 284461 [eHEROES]. T.R. gratefully acknowledges the JungforscherInnenfonds of the Council of the University Graz.  Work at the University of California, Berkeley, was supported from \emph{STEREO} grant NAS5-03131.  The work of K.A., J.R.H., P.C.L. And E.M.D. was conducted at the Jet Propulsion Laboratory, California Institute of Technology under a contract from NASA. N.L. was supported by: AGS-1239704. It is also supported by NASA grant NNX13AP39G and NASA \emph{STEREO} grant to UNH. We acknowledge the use of \WIN data provided by the magnetometer and the solar wind experiment teams at NASA/GSFC, and thank the center for geomagnetism in Kyoto for providing the Dst indices. We also thank the "International study of earth affecting transients" (ISEST) team lead by Jie Zhang.  

\newpage

\newpage





\begin{thebibliography}{87}
\expandafter\ifx\csname natexlab\endcsname\relax\def\natexlab#1{#1}\fi



\bibitem[{{Al-Haddad} {et~al.}(2013){Al-Haddad}, {Nieves-Chinchilla}, {Savani},
  {M{\"o}stl}, {Marubashi}, {Hidalgo}, {Roussev}, {Poedts}, \&
  {Farrugia}}]{alh13}
{Al-Haddad}, N., {et~al.} 2013, \solphys, 284, 129

\bibitem[{{Baker} {et~al.}(2013){Baker}, {Poh}, {Odstrcil}, {Arge}, {Benna},
  {Johnson}, {Korth}, {Gershman}, {Ho}, {McClintock}, {Cassidy}, {Merkel},
  {Raines}, {Schriver}, {Slavin}, {Solomon}, {Tr{\'a}Vn{\'{\i}}{\v c}Ek},
  {Winslow}, \& {Zurbuchen}}]{bak13}
{Baker}, D.~N., {et~al.} 2013, Journal of Geophysical Research (Space Physics),
  118, 45

\bibitem[{{Bothmer} \& {Schwenn}(1998)}]{bot98}
{Bothmer}, V., \& {Schwenn}, R. 1998, Annales Geophysicae, 16, 1

\bibitem[{{Brueckner} {et~al.}(1995){Brueckner}, {Howard}, {Koomen},
  {Korendyke}, {Michels}, {Moses}, {Socker}, {Dere}, {Lamy}, {Llebaria},
  {Bout}, {Schwenn}, {Simnett}, {Bedford}, \& {Eyles}}]{bru95}
{Brueckner}, G.~E., {et~al.} 1995, \solphys, 162, 357

\bibitem[{{Brueckner} {et~al.}(1998){Brueckner}, {Delaboudiniere}, {Howard},
  {Paswaters}, {St.~Cyr}, {Schwenn}, {Lamy}, {Simnett}, {Thompson}, \&
  {Wang}}]{bru98}
---. 1998, \grl, 25, 3019

\bibitem[{{Burlaga} {et~al.}(1981){Burlaga}, {Sittler}, {Mariani}, \&
  {Schwenn}}]{bur81}
{Burlaga}, L., {Sittler}, E., {Mariani}, F., \& {Schwenn}, R. 1981, \jgr, 86,
  6673

\bibitem[{{Colaninno} {et~al.}(2013){Colaninno}, {Vourlidas}, \& {Wu}}]{col13}
{Colaninno}, R.~C., {Vourlidas}, A., \& {Wu}, C.~C. 2013, Journal of
  Geophysical Research (Space Physics), 118, 6866

\bibitem[{{Davies} {et~al.}(2013){Davies}, {Perry}, {Trines}, {Harrison},
  {Lugaz}, {M{\"o}stl}, {Liu}, \& {Steed}}]{dav13}
{Davies}, J.~A., {Perry}, C.~H., {Trines}, R.~M.~G.~M., {Harrison}, R.~A.,
  {Lugaz}, N., {M{\"o}stl}, C., {Liu}, Y.~D., \& {Steed}, K. 2013, \apj, 777,
  167

\bibitem[{{Davies} {et~al.}(2012){Davies}, {Harrison}, {Perry}, {M{\"o}stl},
  {Lugaz}, {Rollett}, {Davis}, {Crothers}, {Temmer}, {Eyles}, \&
  {Savani}}]{dav12}
{Davies}, J.~A., {et~al.} 2012, \apj, 750, 23

\bibitem[{{Davis} {et~al.}(2009){Davis}, {Davies}, {Lockwood}, {Rouillard},
  {Eyles}, \& {Harrison}}]{dav09}
{Davis}, C.~J., {Davies}, J.~A., {Lockwood}, M., {Rouillard}, A.~P., {Eyles},
  C.~J., \& {Harrison}, R.~A. 2009, \grl, 36, 8102

\bibitem[{{Davis} {et~al.}(2011){Davis}, {de Koning}, {Davies}, {Biesecker},
  {Millward}, {Dryer}, {Deehr}, {Webb}, {Schenk}, {Freeland}, {M{\"o}stl},
  {Farrugia}, \& {Odstrcil}}]{dav11}
{Davis}, C.~J., {et~al.} 2011, Space Weather, 9, 1005

\bibitem[{{Galvin} {et~al.}(2008){Galvin}, {Kistler}, {Popecki}, {Farrugia},
  {Simunac}, {Ellis}, {M{\"o}bius}, {Lee}, {Boehm}, {Carroll}, {Crawshaw},
  {Conti}, {Demaine}, {Ellis}, {Gaidos}, {Googins}, {Granoff}, {Gustafson},
  {Heirtzler}, {King}, {Knauss}, {Levasseur}, {Longworth}, {Singer}, {Turco},
  {Vachon}, {Vosbury}, {Widholm}, {Blush}, {Karrer}, {Bochsler}, {Daoudi},
  {Etter}, {Fischer}, {Jost}, {Opitz}, {Sigrist}, {Wurz}, {Klecker}, {Ertl},
  {Seidenschwang}, {Wimmer-Schweingruber}, {Koeten}, {Thompson}, \&
  {Steinfeld}}]{gal08}
{Galvin}, A.~B., {et~al.} 2008, Space Science Reviews, 5

\bibitem[{{Gopalswamy} {et~al.}(2001){Gopalswamy}, {Lara}, {Yashiro}, {Kaiser},
  \& {Howard}}]{gop01}
{Gopalswamy}, N., {Lara}, A., {Yashiro}, S., {Kaiser}, M.~L., \& {Howard},
  R.~A. 2001, \jgr, 106, 29207

\bibitem[{{Gopalswamy} {et~al.}(2009){Gopalswamy}, {M{\"a}kel{\"a}}, {Xie},
  {Akiyama}, \& {Yashiro}}]{gop09}
{Gopalswamy}, N., {M{\"a}kel{\"a}}, P., {Xie}, H., {Akiyama}, S., \& {Yashiro},
  S. 2009, Journal of Geophysical Research (Space Physics), 114, 0

\bibitem[{{Gopalswamy} {et~al.}(2013){Gopalswamy}, {M{\"a}kel{\"a}}, {Xie}, \&
  {Yashiro}}]{gop13}
{Gopalswamy}, N., {M{\"a}kel{\"a}}, P., {Xie}, H., \& {Yashiro}, S. 2013, Space
  Weather, 11, 661

\bibitem[{{Harrison} {et~al.}(2012){Harrison}, {Davies}, {M{\"o}stl}, {Liu},
  {Temmer}, {Bisi}, {Eastwood}, {de Koning}, {Nitta}, {Rollett}, {Farrugia},
  {Forsyth}, {Jackson}, {Jensen}, {Kilpua}, {Odstrcil}, \& {Webb}}]{har12}
{Harrison}, R.~A., {et~al.} 2012, \apj, 750, 45

\bibitem[{{Howard} \& {DeForest}(2012{\natexlab{a}})}]{how12}
{Howard}, T.~A., \& {DeForest}, C.~E. 2012{\natexlab{a}}, \apj, 746, 64

\bibitem[{{Howard} \& {DeForest}(2012{\natexlab{b}})}]{how12b}
---. 2012{\natexlab{b}}, \apj, 752, 130

\bibitem[{{Illing} \& {Hundhausen}(1985)}]{ill85}
{Illing}, R.~M.~E., \& {Hundhausen}, A.~J. 1985, \jgr, 90, 275

\bibitem[{{Isavnin} {et~al.}(2013){Isavnin}, {Vourlidas}, \& {Kilpua}}]{isa13}
{Isavnin}, A., {Vourlidas}, A., \& {Kilpua}, E.~K.~J. 2013, \solphys, 284, 203

\bibitem[{{Janvier} {et~al.}(2013){Janvier}, {D{\'e}moulin}, \&
  {Dasso}}]{jan13}
{Janvier}, M., {D{\'e}moulin}, P., \& {Dasso}, S. 2013, \aap, 556, A50

\bibitem[{{Kaiser} {et~al.}(2008){Kaiser}, {Kucera}, {Davila}, {St.~Cyr},
  {Guhathakurta}, \& {Christian}}]{kai08}
{Kaiser}, M.~L., {Kucera}, T.~A., {Davila}, J.~M., {St.~Cyr}, O.~C.,
  {Guhathakurta}, M., \& {Christian}, E. 2008, Space Science Reviews, 136, 5

\bibitem[{{Kilpua} {et~al.}(2012){Kilpua}, {Mierla}, {Rodriguez}, {Zhukov},
  {Srivastava}, \& {West}}]{kil12b}
{Kilpua}, E.~K.~J., {Mierla}, M., {Rodriguez}, L., {Zhukov}, A.~N.,
  {Srivastava}, N., \& {West}, M.~J. 2012, \solphys, 279, 477

\bibitem[{{Kilpua} {et~al.}(2009){Kilpua}, {Pomoell}, {Vourlidas}, {Vainio},
  {Luhmann}, {Li}, {Schroeder}, {Galvin}, \& {Simunac}}]{kil09c}
{Kilpua}, E.~K.~J., {et~al.} 2009, Annales Geophysicae, 27, 4491

\bibitem[{{Leitner} {et~al.}(2007){Leitner}, {Farrugia}, {M{\"o}stl},
  {Ogilvie}, {Galvin}, {Schwenn}, \& {Biernat}}]{lei07}
{Leitner}, M., {Farrugia}, C.~J., {M{\"o}stl}, C., {Ogilvie}, K.~W., {Galvin},
  A.~B., {Schwenn}, R., \& {Biernat}, H.~K. 2007, Journal of Geophysical
  Research (Space Physics), 112, 6113

\bibitem[{{Lepping} {et~al.}(1995){Lepping}, {Acuna}, {Burlaga}, {Farrell},
  {Slavin}, {Schatten}, {Mariani}, {Ness}, {Neubauer}, {Whang}, {Byrnes},
  {Kennon}, {Panetta}, {Scheifele}, \& {Worley}}]{lep95}
{Lepping}, R.~P., {et~al.} 1995, Space Science Reviews, 71, 207

\bibitem[{{Liewer} {et~al.}(2008){Liewer}, {Ayon}, {Alexander}, {Kosovichev},
  {Mewaldt}, {Socker}, \& {Vourlidas}}]{lie08}
{Liewer}, P.~C., {Ayon}, J., {Alexander}, D., {Kosovichev}, A., {Mewaldt},
  R.~A., {Socker}, D.~G., \& {Vourlidas}, A. 2008, {Solar Polar Imager:
  Observing Solar Activity from a New Perspective} (American Institute of
  Aeronautics and Astronautics), 1--56347

\bibitem[{{Liewer} {et~al.}(2011){Liewer}, {Hall}, {Howard}, {De Jong},
  {Thompson}, \& {Thernisien}}]{lie11}
{Liewer}, P.~C., {Hall}, J.~R., {Howard}, R.~A., {De Jong}, E.~M., {Thompson},
  W.~T., \& {Thernisien}, A. 2011, Journal of Atmospheric and Solar-Terrestrial
  Physics, 73, 1173

\bibitem[{{Lindsay} {et~al.}(1999){Lindsay}, {Luhmann}, {Russell}, \&
  {Gosling}}]{lin99}
{Lindsay}, G.~M., {Luhmann}, J.~G., {Russell}, C.~T., \& {Gosling}, J.~T. 1999,
  \jgr, 104, 12515

\bibitem[{{Liu} {et~al.}(2010{\natexlab{a}}){Liu}, {Davies}, {Luhmann},
  {Vourlidas}, {Bale}, \& {Lin}}]{liu10}
{Liu}, Y., {Davies}, J.~A., {Luhmann}, J.~G., {Vourlidas}, A., {Bale}, S.~D.,
  \& {Lin}, R.~P. 2010{\natexlab{a}}, \apjl, 710, L82

\bibitem[{{Liu} {et~al.}(2011){Liu}, {Luhmann}, {Bale}, \& {Lin}}]{liu11}
{Liu}, Y., {Luhmann}, J.~G., {Bale}, S.~D., \& {Lin}, R.~P. 2011, \apj, 734, 84

\bibitem[{{Liu} {et~al.}(2008){Liu}, {Luhmann}, {Huttunen}, {Lin}, {Bale},
  {Russell}, \& {Galvin}}]{liu08}
{Liu}, Y., {Luhmann}, J.~G., {Huttunen}, K.~E.~J., {Lin}, R.~P., {Bale}, S.~D.,
  {Russell}, C.~T., \& {Galvin}, A.~B. 2008, \apjl, 677, L133

\bibitem[{{Liu} {et~al.}(2005){Liu}, {Richardson}, \& {Belcher}}]{liu05}
{Liu}, Y., {Richardson}, J.~D., \& {Belcher}, J.~W. 2005, \planss, 53, 3

\bibitem[{{Liu} {et~al.}(2010{\natexlab{b}}){Liu}, {Thernisien}, {Luhmann},
  {Vourlidas}, {Davies}, {Lin}, \& {Bale}}]{liu10b}
{Liu}, Y., {Thernisien}, A., {Luhmann}, J.~G., {Vourlidas}, A., {Davies},
  J.~A., {Lin}, R.~P., \& {Bale}, S.~D. 2010{\natexlab{b}}, \apj, 722, 1762

\bibitem[{{Liu} {et~al.}(2013){Liu}, {Luhmann}, {Lugaz}, {M{\"o}stl}, {Davies},
  {Bale}, \& {Lin}}]{liu13}
{Liu}, Y.~D., {Luhmann}, J.~G., {Lugaz}, N., {M{\"o}stl}, C., {Davies}, J.~A.,
  {Bale}, S.~D., \& {Lin}, R.~P. 2013, \apj, 769, 45

\bibitem[{{Liu} {et~al.}(2012){Liu}, {Luhmann}, {M{\"o}stl},
  {Martinez-Oliveros}, {Bale}, {Lin}, {Harrison}, {Temmer}, {Webb}, \&
  {Odstrcil}}]{liu12}
{Liu}, Y.~D., {et~al.} 2012, \apjl, 746, L15

\bibitem[{{Lugaz}(2010)}]{lug10b}
{Lugaz}, N. 2010, \solphys, 267, 411

\bibitem[{{Lugaz} {et~al.}(2010){Lugaz}, {Hernandez-Charpak}, {Roussev},
  {Davis}, {Vourlidas}, \& {Davies}}]{lug10}
{Lugaz}, N., {Hernandez-Charpak}, J.~N., {Roussev}, I.~I., {Davis}, C.~J.,
  {Vourlidas}, A., \& {Davies}, J.~A. 2010, \apj, 715, 493

\bibitem[{{Lugaz} \& {Kintner}(2013)}]{lug13}
{Lugaz}, N., \& {Kintner}, P. 2013, \solphys, 285, 281

\bibitem[{{Lugaz} {et~al.}(2012){Lugaz}, {Kintner}, {M{\"o}stl}, {Jian},
  {Davis}, \& {Farrugia}}]{lug12}
{Lugaz}, N., {Kintner}, P., {M{\"o}stl}, C., {Jian}, L.~K., {Davis}, C.~J., \&
  {Farrugia}, C.~J. 2012, \solphys, 279, 497

\bibitem[{{Lugaz} {et~al.}(2011){Lugaz}, {Roussev}, \& {Gombosi}}]{lug11}
{Lugaz}, N., {Roussev}, I.~I., \& {Gombosi}, T.~I. 2011, Advances in Space
  Research, 48, 292

\bibitem[{{Lugaz} {et~al.}(2009){Lugaz}, {Vourlidas}, \& {Roussev}}]{lug09a}
{Lugaz}, N., {Vourlidas}, A., \& {Roussev}, I.~I. 2009, Annales Geophysicae,
  27, 3479

\bibitem[{{Luhmann} {et~al.}(2008){Luhmann}, {Curtis}, {Schroeder}, {McCauley},
  {Lin}, {Larson}, {Bale}, {Sauvaud}, {Aoustin}, {Mewaldt}, {Cummings},
  {Stone}, {Davis}, {Cook}, {Kecman}, {Wiedenbeck}, {von Rosenvinge}, {Acuna},
  {Reichenthal}, {Shuman}, {Wortman}, {Reames}, {Mueller-Mellin}, {Kunow},
  {Mason}, {Walpole}, {Korth}, {Sanderson}, {Russell}, \& {Gosling}}]{luh08}
{Luhmann}, J.~G., {et~al.} 2008, Space Science Reviews, 136, 117

\bibitem[{{Lynch} {et~al.}(2010){Lynch}, {Li}, {Thernisien}, {Robbrecht},
  {Fisher}, {Luhmann}, \& {Vourlidas}}]{lyn10}
{Lynch}, B.~J., {Li}, Y., {Thernisien}, A.~F.~R., {Robbrecht}, E., {Fisher},
  G.~H., {Luhmann}, J.~G., \& {Vourlidas}, A. 2010, Journal of Geophysical
  Research (Space Physics), 115, 7106

\bibitem[{{Lynch} {et~al.}(2003){Lynch}, {Zurbuchen}, {Fisk}, \&
  {Antiochos}}]{lyn03}
{Lynch}, B.~J., {Zurbuchen}, T.~H., {Fisk}, L.~A., \& {Antiochos}, S.~K. 2003,
  Journal of Geophysical Research (Space Physics), 108, 1239

\bibitem[{{Mishra} \& {Srivastava}(2013)}]{mis13}
{Mishra}, W., \& {Srivastava}, N. 2013, \apj, 772, 70

\bibitem[{{Mishra} {et~al.}(2014){Mishra}, {Srivastava}, \& {Davies}}]{mis14}
{Mishra}, W., {Srivastava}, N., \& {Davies}, J.~A. 2014, \apj, 784, 135

\bibitem[{{M{\"o}stl} \& {Davies}(2013)}]{moe13}
{M{\"o}stl}, C., \& {Davies}, J.~A. 2013, \solphys, 285, 411

\bibitem[{{M{\"o}stl} {et~al.}(2009{\natexlab{a}}){M{\"o}stl}, {Farrugia},
  {Biernat}, {Leitner}, {Kilpua}, {Galvin}, \& {Luhmann}}]{moe09b}
{M{\"o}stl}, C., {Farrugia}, C.~J., {Biernat}, H.~K., {Leitner}, M., {Kilpua},
  E.~K.~J., {Galvin}, A.~B., \& {Luhmann}, J.~G. 2009{\natexlab{a}}, \solphys,
  256, 427

\bibitem[{{M{\"o}stl} {et~al.}(2009{\natexlab{b}}){M{\"o}stl}, {Farrugia},
  {Temmer}, {Miklenic}, {Veronig}, {Galvin}, {Leitner}, \& {Biernat}}]{moe09c}
{M{\"o}stl}, C., {Farrugia}, C.~J., {Temmer}, M., {Miklenic}, C., {Veronig},
  A.~M., {Galvin}, A.~B., {Leitner}, M., \& {Biernat}, H.~K.
  2009{\natexlab{b}}, \apjl, 705, L180

\bibitem[{{M{\"o}stl} {et~al.}(2009{\natexlab{c}}){M{\"o}stl}, {Farrugia},
  {Miklenic}, {Temmer}, {Galvin}, {Luhmann}, {Kilpua}, {Leitner},
  {Nieves-Chinchilla}, {Veronig}, \& {Biernat}}]{moe09}
{M{\"o}stl}, C., {et~al.} 2009{\natexlab{c}}, Journal of Geophysical Research
  (Space Physics), 114, 4102

\bibitem[{{M{\"o}stl} {et~al.}(2010){M{\"o}stl}, {Temmer}, {Rollett},
  {Farrugia}, {Liu}, {Veronig}, {Leitner}, {Galvin}, \& {Biernat}}]{moe10}
---. 2010, \grl, 37, L24103

\bibitem[{{M{\"o}stl} {et~al.}(2011){M{\"o}stl}, {Rollett}, {Lugaz},
  {Farrugia}, {Davies}, {Temmer}, {Veronig}, {Harrison}, {Crothers}, {Luhmann},
  {Galvin}, {Zhang}, {Baumjohann}, \& {Biernat}}]{moe11}
---. 2011, \apj, 741, 34

\bibitem[{{M{\"o}stl} {et~al.}(2012){M{\"o}stl}, {Farrugia}, {Kilpua}, {Jian},
  {Liu}, {Eastwood}, {Harrison}, {Webb}, {Temmer}, {Odstrcil}, {Davies},
  {Rollett}, {Luhmann}, {Nitta}, {Mulligan}, {Jensen}, {Forsyth}, {Lavraud},
  {de Koning}, {Veronig}, {Galvin}, {Zhang}, \& {Anderson}}]{moe12}
---. 2012, \apj, 758, 10

\bibitem[{{M{\"u}ller} {et~al.}(2013){M{\"u}ller}, {Marsden}, {St.~Cyr}, \&
  {Gilbert}}]{mue13}
{M{\"u}ller}, D., {Marsden}, R.~G., {St.~Cyr}, O.~C., \& {Gilbert}, H.~R. 2013,
  \solphys, 285, 25

\bibitem[{{Mulligan} {et~al.}(1998){Mulligan}, {Russell}, \& {Luhmann}}]{mul98}
{Mulligan}, T., {Russell}, C.~T., \& {Luhmann}, J.~G. 1998, \grl, 25, 2959

\bibitem[{{Ogilvie} {et~al.}(1995){Ogilvie}, {Chornay}, {Fritzenreiter},
  {Hunsaker}, {Keller}, {Lobell}, {Miller}, {Scudder}, {Sittler}, {Torbert},
  {Bodet}, {Needell}, {Lazarus}, {Steinberg}, {Tappan}, {Mavretic}, \&
  {Gergin}}]{ogi95}
{Ogilvie}, K.~W., {et~al.} 1995, Space Science Reviews, 71, 55

\bibitem[{{Qiu} {et~al.}(2007){Qiu}, {Hu}, {Howard}, \& {Yurchyshyn}}]{qiu07}
{Qiu}, J., {Hu}, Q., {Howard}, T.~A., \& {Yurchyshyn}, V.~B. 2007, \apj, 659,
  758

\bibitem[{{Qiu} \& {Yurchyshyn}(2005)}]{qiu05}
{Qiu}, J., \& {Yurchyshyn}, V.~B. 2005, \apjl, 634, L121

\bibitem[{{Richardson} \& {Cane}(2010)}]{ric10}
{Richardson}, I.~G., \& {Cane}, H.~V. 2010, \solphys, 264, 189

\bibitem[{{Rollett} {et~al.}(2012){Rollett}, {M{\"o}stl}, {Temmer}, {Veronig},
  {Farrugia}, \& {Biernat}}]{rol12}
{Rollett}, T., {M{\"o}stl}, C., {Temmer}, M., {Veronig}, A.~M., {Farrugia},
  C.~J., \& {Biernat}, H.~K. 2012, \solphys, 276, 293

\bibitem[{{Rollett} {et~al.}(2013){Rollett}, {Temmer}, {M{\"o}stl}, {Lugaz},
  {Veronig}, \& {M{\"o}stl}}]{rol13}
{Rollett}, T., {Temmer}, M., {M{\"o}stl}, C., {Lugaz}, N., {Veronig}, A.~M., \&
  {M{\"o}stl}, U.~V. 2013, \solphys, 283, 541

\bibitem[{{Rouillard}(2011)}]{rou11rev}
{Rouillard}, A.~P. 2011, Journal of Atmospheric and Solar-Terrestrial Physics,
  73, 1201

\bibitem[{{Rouillard} {et~al.}(2008){Rouillard}, {Davies}, {Forsyth}, {Rees},
  {Davis}, {Harrison}, {Lockwood}, {Bewsher}, {Crothers}, {Eyles}, {Hapgood},
  \& {Perry}}]{rou08}
{Rouillard}, A.~P., {et~al.} 2008, \grl, 35, 10110

\bibitem[{{Rouillard} {et~al.}(2009){Rouillard}, {Davies}, {Forsyth}, {Savani},
  {Sheeley}, {Thernisien}, {Zhang}, {Howard}, {Anderson}, {Carr}, {Tsang},
  {Lockwood}, {Davis}, {Harrison}, {Bewsher}, {Fr{\"a}nz}, {Crothers}, {Eyles},
  {Brown}, {Whittaker}, {Hapgood}, {Coates}, {Jones}, {Grande}, {Frahm}, \&
  {Winningham}}]{rou09}
---. 2009, Journal of Geophysical Research (Space Physics), 114, 7106

\bibitem[{{Savani} {et~al.}(2010){Savani}, {Owens}, {Rouillard}, {Forsyth}, \&
  {Davies}}]{sav10}
{Savani}, N.~P., {Owens}, M.~J., {Rouillard}, A.~P., {Forsyth}, R.~J., \&
  {Davies}, J.~A. 2010, \apjl, 714, L128

\bibitem[{{Savani} {et~al.}(2012){Savani}, {Davies}, {Davis}, {Shiota},
  {Rouillard}, {Owens}, {Kusano}, {Bothmer}, {Bamford}, {Lintott}, \&
  {Smith}}]{sav12b}
{Savani}, N.~P., {et~al.} 2012, \solphys, 279, 517

\bibitem[{{Schwenn} {et~al.}(2005){Schwenn}, {dal Lago}, {Huttunen}, \&
  {Gonzalez}}]{sch05}
{Schwenn}, R., {dal Lago}, A., {Huttunen}, E., \& {Gonzalez}, W.~D. 2005,
  Annales Geophysicae, 23, 1033

\bibitem[{{Sheeley} {et~al.}(1999){Sheeley}, {Walters}, {Wang}, \&
  {Howard}}]{she99}
{Sheeley}, N.~R., {Walters}, J.~H., {Wang}, Y., \& {Howard}, R.~A. 1999, \jgr,
  104, 24739

\bibitem[{{Temmer} {et~al.}(2011){Temmer}, {Rollett}, {M{\"o}stl}, {Veronig},
  {Vr{\v s}nak}, \& {Odstr{\v c}il}}]{tem11}
{Temmer}, M., {Rollett}, T., {M{\"o}stl}, C., {Veronig}, A.~M., {Vr{\v s}nak},
  B., \& {Odstr{\v c}il}, D. 2011, \apj, 743, 101

\bibitem[{{Temmer} {et~al.}(2012){Temmer}, {Vr{\v s}nak}, {Rollett}, {Bein},
  {de Koning}, {Liu}, {Bosman}, {Davies}, {M{\"o}stl}, {{\v Z}ic}, {Veronig},
  {Bothmer}, {Harrison}, {Nitta}, {Bisi}, {Flor}, {Eastwood}, {Odstrcil}, \&
  {Forsyth}}]{tem12}
{Temmer}, M., {et~al.} 2012, \apj, 749, 57

\bibitem[{{Thernisien} {et~al.}(2009){Thernisien}, {Vourlidas}, \&
  {Howard}}]{the09}
{Thernisien}, A., {Vourlidas}, A., \& {Howard}, R.~A. 2009, \solphys, 256, 111

\bibitem[{{Thernisien} {et~al.}(2006){Thernisien}, {Howard}, \&
  {Vourlidas}}]{the06}
{Thernisien}, A.~F.~R., {Howard}, R.~A., \& {Vourlidas}, A. 2006, \apj, 652,
  763

\bibitem[{{Vourlidas} {et~al.}(2011){Vourlidas}, {Colaninno},
  {Nieves-Chinchilla}, \& {Stenborg}}]{vou11}
{Vourlidas}, A., {Colaninno}, R., {Nieves-Chinchilla}, T., \& {Stenborg}, G.
  2011, \apjl, 733, L23

\bibitem[{{Vourlidas} \& {Howard}(2006)}]{vou06}
{Vourlidas}, A., \& {Howard}, R.~A. 2006, \apj, 642, 1216

\bibitem[{{Vourlidas} {et~al.}(2013){Vourlidas}, {Lynch}, {Howard}, \&
  {Li}}]{vou13}
{Vourlidas}, A., {Lynch}, B.~J., {Howard}, R.~A., \& {Li}, Y. 2013, \solphys,
  284, 179

\bibitem[{{Vr{\v s}nak} \& {{\v Z}ic}(2007)}]{vrs07}
{Vr{\v s}nak}, B., \& {{\v Z}ic}, T. 2007, \aap, 472, 937

\bibitem[{{Vr{\v s}nak} {et~al.}(2013){Vr{\v s}nak}, {{\v Z}ic}, {Vrbanec},
  {Temmer}, {Rollett}, {M{\"o}stl}, {Veronig}, {{\v C}alogovi{\'c}},
  {Dumbovi{\'c}}, {Luli{\'c}}, {Moon}, \& {Shanmugaraju}}]{vrs13}
{Vr{\v s}nak}, B., {et~al.} 2013, \solphys, 285, 295

\bibitem[{{Webb} {et~al.}(2013){Webb}, {M{\"o}stl}, {Jackson}, {Bisi},
  {Howard}, {Mulligan}, {Jensen}, {Jian}, {Davies}, {de Koning}, {Liu},
  {Temmer}, {Clover}, {Farrugia}, {Harrison}, {Nitta}, {Odstrcil}, {Tappin}, \&
  {Yu}}]{web13}
{Webb}, D.~F., {et~al.} 2013, \solphys, 285, 317

\bibitem[{{Williams} {et~al.}(2009){Williams}, {Davies}, {Milan}, {Rouillard},
  {Davis}, {Perry}, \& {Harrison}}]{wil09}
{Williams}, A.~O., {Davies}, J.~A., {Milan}, S.~E., {Rouillard}, A.~P.,
  {Davis}, C.~J., {Perry}, C.~H., \& {Harrison}, R.~A. 2009, Annales
  Geophysicae, 27, 4359

\bibitem[{{Wood} {et~al.}(2010){Wood}, {Howard}, \& {Socker}}]{woo10}
{Wood}, B.~E., {Howard}, R.~A., \& {Socker}, D.~G. 2010, \apj, 715, 1524

\bibitem[{{Xiong} {et~al.}(2013{\natexlab{a}}){Xiong}, {Davies}, {Bisi},
  {Owens}, {Fallows}, \& {Dorrian}}]{xio13b}
{Xiong}, M., {Davies}, J.~A., {Bisi}, M.~M., {Owens}, M.~J., {Fallows}, R.~A.,
  \& {Dorrian}, G.~D. 2013{\natexlab{a}}, \solphys, 285, 369

\bibitem[{{Xiong} {et~al.}(2013{\natexlab{b}}){Xiong}, {Davies}, {Feng},
  {Owens}, {Harrison}, {Davis}, \& {Liu}}]{xio13}
{Xiong}, M., {Davies}, J.~A., {Feng}, X., {Owens}, M.~J., {Harrison}, R.~A.,
  {Davis}, C.~J., \& {Liu}, Y.~D. 2013{\natexlab{b}}, \apj, 777, 32

\bibitem[{{Yurchyshyn} {et~al.}(2005){Yurchyshyn}, {Hu}, \&
  {Abramenko}}]{yur05}
{Yurchyshyn}, V., {Hu}, Q., \& {Abramenko}, V. 2005, Space Weather, 3, 8

\bibitem[{{Zhang} {et~al.}(2007){Zhang}, {Richardson}, {Webb}, {Gopalswamy},
  {Huttunen}, {Kasper}, {Nitta}, {Poomvises}, {Thompson}, {Wu}, {Yashiro}, \&
  {Zhukov}}]{zha07}
{Zhang}, J., {et~al.} 2007, Journal of Geophysical Research (Space Physics),
  112, 10102

\end{thebibliography}

\clearpage




\clearpage

\clearpage






\end{document}